\documentclass[preprint,12pt]{elsarticle}

\usepackage{amssymb}
\usepackage{amsmath}
\usepackage{url}
\usepackage{mathrsfs}
\usepackage{graphicx}%
\usepackage{multirow}%
\usepackage{amsmath,amssymb,amsfonts}%
\usepackage{xcolor}%
\usepackage{subcaption}
\usepackage{tikz}
\usetikzlibrary{arrows.meta, positioning, shapes.geometric}
\usepackage{textcomp}%
\usepackage{manyfoot}%
\usepackage{booktabs}%
\usepackage{algorithm}%
\usepackage{algorithmicx}%
\usepackage{algpseudocode}%
\usepackage{listings}%
\usepackage{graphicx}
\usepackage{svg} 
\usepackage{caption}   
\usepackage{hyperref}

\journal{Advanced Engineering Informatics}

\begin{document}

\begin{frontmatter}






\title{Robust Short-Term OEE Forecasting in Industry~4.0 via Topological Data Analysis}

\author[1]{Korkut Anapa}
\ead{korkut.anapa@metu.edu.tr}

\author[1,2]{İsmail Güzel}

\ead{iguzel@metu.edu.tr}

\author[3]{Ceylan Yozgatlıgil}
\ead{ceylan@metu.edu.tr}

\affiliation[1]{organization={Institute of Applied Mathematics, Middle East Technical University},
    addressline={Üniversiteler Mahallesi, Dumlupınar Bulvarı No:1},
    city={Ankara},
    postcode={06800},
    country={Türkiye}}

\affiliation[2]{organization={Network Technologies Department, TÜBİTAK ULAKBİM},
    addressline={Üniversiteler Mahallesi, Dumlupınar Bulvarı No:1},
    city={Ankara},
    postcode={06800},
    country={Türkiye}}

\affiliation[3]{organization={Department of Statistics, Middle East Technical University},
    addressline={Üniversiteler Mahallesi, Dumlupınar Bulvarı No:1},
    city={Ankara},
    postcode={06800},
    country={Türkiye}}

\begin{abstract}
In Industry 4.0 manufacturing environments, forecasting Overall Equipment Efficiency (OEE) is critical for data-driven operational control and predictive maintenance. However, the highly volatile and nonlinear nature of OEE time series—particularly in complex production lines and hydraulic press systems—limits the effectiveness of forecasting. This study proposes a novel informational framework that leverages Topological Data Analysis (TDA) to transform raw OEE data into structured engineering knowledge for production management.

The framework models hourly OEE data from production lines and  systems using persistent homology to extract large-scale topological features that characterize intrinsic operational behaviors. These features are integrated into a SARIMAX (Seasonal Autoregressive Integrated Moving Average with Exogenous Regressors) architecture, where TDA components serve as exogenous variables to capture latent temporal structures. Experimental results demonstrate forecasting accuracy improvements of at least 17\% over standard seasonal benchmarks, with Heat Kernel–based features consistently identified as the most effective predictors.

The proposed framework was deployed in a Global Lighthouse Network manufacturing facility, providing a new strategic layer for production management and achieving a 7.4\% improvement in total OEE. This research contributes a formal methodology for embedding topological signatures into classical stochastic models to enhance decision-making in knowledge-intensive production systems.
\end{abstract}


\begin{graphicalabstract}
  \centering
  \includegraphics[width=\linewidth]{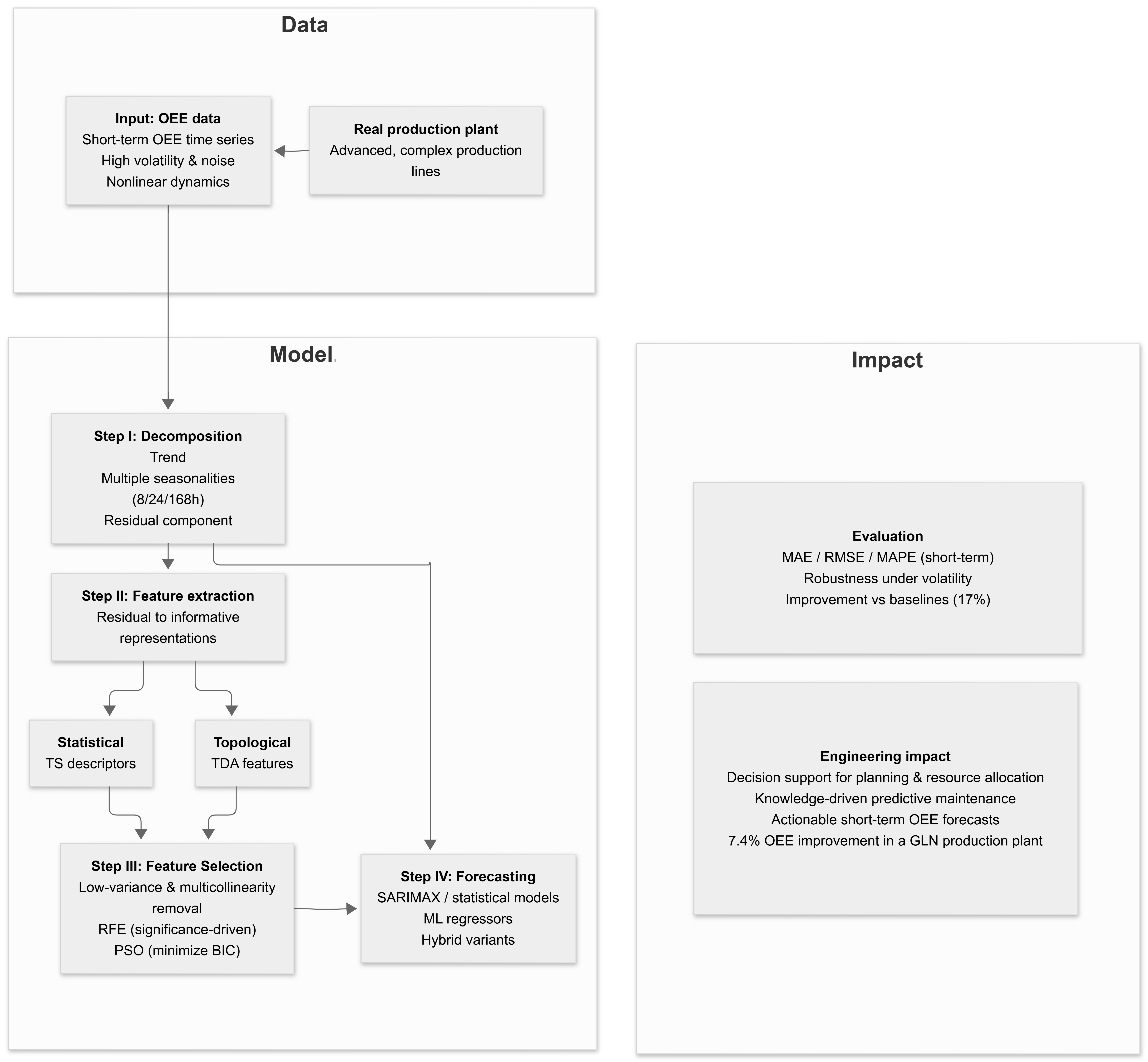}
\end{graphicalabstract}

\begin{highlights}
\item A Topological Data Analysis (TDA) based framework for managing production systems.

\item TDA enhances residual-based decomposition for robust short-term OEE forecasting.

\item Persistent homology-derived features improve SARIMAX forecasting accuracy by 17\%.

\item Heat Kernel based features are superior predictors for OEE volatility.

\item Deployment in a Global Lighthouse facility achieved 7.4\% OEE improvement.

\end{highlights}


\end{frontmatter}



\section{Introduction}\label{sec1}

In contemporary manufacturing systems, performance improvements increasingly arise not from isolated technological upgrades but from the ability to extract, represent, and reason over operational knowledge embedded in production data. As production environments become more complex and data-rich, engineering challenges shift toward knowledge-intensive tasks that require advanced computational representations and decision-support mechanisms like proactive maintenance and dynamic production scheduling. Within this context, engineering informatics plays a central role by enabling the transformation of raw operational measurements into structured knowledge that can support informed engineering decisions.

Key performance indicators such as Overall Equipment Efficiency (OEE) provide concise summaries of production behavior and are widely used for operational monitoring. However, in complex machinery such as stainless-steel tubs production lines and hydraulic press systems, while OEE is straightforward to compute, the underlying temporal dynamics are often highly nonlinear, volatile, and difficult to interpret using conventional statistical descriptors. By leveraging advanced analytical strategies such as Topological Data Analysis (TDA), it becomes possible to extract structured representations that encode the intrinsic geometric and dynamical properties of OEE time series. Integrating these topological signatures into hybrid forecasting architectures, such as SARIMAX, allows for the capturing of latent operational states. These representations enable reasoning about system behaviour, thereby supporting smarter, more proactive production management and predictive maintenance within Industry 4.0 manufacturing environments.

\subsection{Motivation}\label{subsec2}

OEE is a widely adopted indicator for evaluating operational performance in manufacturing systems \cite{1}. Introduced by Nakajima \cite{2}, OEE integrates availability, performance, and quality into a single metric and has been extensively used to monitor machines, manufacturing cells, and assembly lines, supporting productivity analysis and continuous improvement initiatives \cite{3,4,5}. Due to its simplicity and accessibility, OEE remains a central reference for both operators and managers in guiding maintenance strategies and improving overall equipment effectiveness \cite{6,7}.

In Industry~4.0 manufacturing environments, OEE evolves from a static performance indicator into a highly dynamic time series influenced by complex interactions among machines, operators, materials, and control policies. High-frequency monitoring and increasing system complexity result in OEE signals that exhibit strong volatility, nonlinear dynamics, and abrupt structural changes. Under these conditions, extracting decision-relevant information from OEE data becomes a knowledge-intensive engineering task that exceeds the descriptive capacity of classical statistical analysis.

A key application of OEE in modern manufacturing is short-term forecasting to assess production system performance and to anticipate potential equipment failures. This capability, together with the adoption of advanced digital technologies, data-driven analytics, and information engineering projects, is increasingly emphasised in World Economic Forum Global Lighthouse Network (GLN) factories, where use cases aligned with Total Productive Maintenance (TPM) and World Class Manufacturing (WCM) practices demonstrate the industrial value of predictive analytics \cite{8}. Although several studies have investigated OEE forecasting \cite{9,10,11,12}, limitations in forecasting accuracy and in the expressive power of existing modeling approaches continue to hinder the development of precise and actionable decision-support frameworks for production systems.

Classical time-series models such as AutoRegressive Integrated Moving Average (ARIMA), exponential smoothing, and seasonal decomposition have been widely applied to manufacturing data to capture trends and recurring patterns \cite{13,14}. While effective in relatively stable conditions, these approaches often struggle when confronted with highly volatile, nonlinear, and structurally complex OEE dynamics. Recent advances in artificial intelligence and machine learning have improved predictive performance in certain scenarios \cite{15,16}, yet many data-driven models remain limited in their ability to explicitly represent and reason about the underlying temporal structure of operational behavior \cite{17,18,19,20,21}. As a result, short-term OEE forecasting remains challenging when forecasted values are expected to function as decision-relevant knowledge.

Motivated by these challenges, this study develops an engineering informatics framework that supports expressive temporal representation and reliable short-term forecasting of OEE under highly dynamic operating conditions. The framework aims to bridge the gap between forecasting accuracy and actionable engineering insight, offering the following key contributions:

\begin{enumerate}
\item \textbf{Knowledge-oriented forecasting framework for volatile OEE dynamics:}
A forecasting framework is proposed for short-term, highly volatile OEE time series, emphasizing structured information extraction. By integrating statistical decomposition with expressive temporal representations, the framework enables robust forecasting under noise, nonstationarity, and limited data availability.

\item \textbf{Topological representation of temporal behavior:}
TDA is introduced to transform OEE time series into structured feature representations that capture intrinsic geometric and topological properties of operational dynamics. These representations remain informative under strong volatility and nonlinear behaviour, providing forecastable characterisations of equipment performance.

\item \textbf{Performance-driven hybrid feature selection strategy:}
A multi-stage feature selection methodology is developed to manage high dimensional statistical and topological feature spaces. The strategy combines SARIMAX-based statistical significance testing, recursive feature elimination, and Particle Swarm Optimisation guided by the Bayesian Information Criterion, ensuring predictive relevance, interpretability, and computational efficiency.

\item \textbf{Industrial validation in a GLN manufacturing facility:}
The proposed framework is deployed in a Global Lighthouse Network production plant, where it supports real-time OEE monitoring and short-term forecasting across multiple assets. The deployment demonstrates the practical applicability and transferability of the proposed approach, yielding measurable improvements in equipment effectiveness.
\end{enumerate}

This study builds upon prior work \cite{22}, which investigated short-term OEE forecasting for hydraulic press systems and highlighted the limitations of traditional statistical and artificial intelligence approaches under highly volatile operating conditions. Extending this foundation, the present work advances the methodological design, conducts a large-scale analysis of topological feature representations, and evaluates the framework across diverse production systems, deepening the understanding of how topological representations can support knowledge-intensive forecasting and decision-making in modern manufacturing environments.

The remainder of this paper is organized as follows. Section~1 reviews related work. Section~2 describes the proposed methodology and its main components. Section~3 presents the feature extraction and feature selection strategies. Section~4 introduces the forecasting models. Section~5 reports the experimental results, Section~6 demonstrates the application of the framework in a real manufacturing environment, and Section~7 discusses the findings.

\subsection{Literature Review}
\label{literature}

\subsubsection{OEE in the Modern Era}
\label{subsec1}

The Fourth Industrial Revolution (4IR) has substantially reshaped OEE measurement and management through the integration of Internet of Things (IoT) infrastructures, machine learning techniques, and advanced data analytics \cite{8,10}. These technologies have enabled high-frequency data acquisition and real-time OEE monitoring \cite{9,11}, as well as predictive analytics frameworks for forecasting efficiency trends and anticipating potential failures \cite{12}. Consequently, OEE management has progressively evolved from reactive performance assessment toward predictive and optimization-oriented strategies that support maintenance planning and resource allocation in modern manufacturing systems \cite{11}.

Despite these advancements, the predictive accuracy of existing OEE analytics remains limited when applied to short-term, highly volatile OEE time series. Most contemporary forecasting approaches are designed to capture smooth trends or aggregated behavior and therefore rely on assumptions such as local stationarity, temporal continuity, or noise-averaging effects \cite{13,14}. However, operational OEE data at fine temporal resolutions are often dominated by abrupt fluctuations arising from micro-stoppages, transient process disturbances, operator interventions, and rapid production regime changes. Under such conditions, traditional descriptors fail to capture the underlying state transitions, and prediction errors tend to increase significantly, reducing the reliability of short-horizon OEE forecasts precisely where timely and accurate estimates are most critical for operational decision-making \cite{15,16}.

Moreover, short-term OEE volatility is frequently treated as stochastic noise rather than as an informative manifestation of underlying system dynamics. From an engineering informatics perspective, this constitutes a loss of critical operational knowledge. This leads to over-smoothed predictions or delayed model responses that fail to track rapid efficiency changes in real time \cite{17}. As a result, existing OEE-related studies predominantly emphasize numerical accuracy over longer horizons or focus on downstream optimization objectives, while the challenge of achieving robust and reliable forecasting performance under highly dynamic operating conditions remains insufficiently addressed \cite{18,20}. This gap highlights the need for engineering informatics approaches capable of capturing the intrinsic temporal structure of volatile OEE signals, thereby improving short-term predictive accuracy and enhancing the operational value of OEE forecasting in Industry~4.0 environments.

\subsubsection{Forecasting}

Forecasting methodologies have evolved from classical statistical models toward increasingly complex data-driven approaches. Traditional time series models such as AutoRegressive Integrated Moving Average (ARIMA) have been widely used due to their mathematical interpretability and efficiency in modeling linear temporal dependencies in stationary data \cite{13}. However, the reliance of ARIMA-based approaches on linear assumptions and carefully tuned differencing limits their applicability when forecasting highly volatile, nonlinear, and nonstationary time series commonly observed in modern engineering systems \cite{14}.

To address these limitations, machine learning and artificial intelligence methods have been extensively explored in forecasting applications. Techniques including support vector regression, random forests, gradient boosting machines, and neural networks—particularly long short-term memory architectures—have demonstrated strong performance in capturing nonlinear dependencies in long and complex time series \cite{23,24,25,26}. The inclusion of exogenous variables and engineered statistical features has further improved predictive accuracy in many applications \cite{14,27}. Despite these advances, many ML-based forecasting models remain largely data-driven and opaque, offering limited interpretability and restricted capability to explicitly represent the underlying temporal structure governing system behavior.

More recently, forecasting research has begun to explore the use of topological representations derived from time series data. TDA, grounded in persistent homology, provides tools for characterizing the geometric and structural properties of dynamical systems beyond pointwise numerical observations \cite{28}. Features derived from persistence diagrams, Betti curves, and persistence entropy have shown promise for capturing intrinsic temporal structure, particularly in short-term, highly volatile datasets where conventional statistical and ML models often struggle. Applications in domains such as financial time series forecasting illustrate the potential of topological features to encode robust and informative representations in environments characterized by noise and uncertainty \cite{29,30}.

Overall, while classical statistical models and modern ML approaches have advanced forecasting performance, their ability to support knowledge-intensive reasoning in complex and volatile systems remains limited \cite{33}. The integration of topological representations offers a complementary perspective by transforming raw temporal data into structured knowledge, providing a foundation for forecasting approaches that emphasize interpretability, robustness, and decision support in engineering applications.

\subsubsection{Topological Data Analysis}

TDA has emerged as a powerful framework for analyzing complex and high-dimensional data, particularly in settings characterized by noise, nonlinearity, and structural variability. A central advantage of TDA lies in its ability to capture intrinsic data structure through topological features that remain stable under perturbations, enabling robust pattern extraction beyond pointwise numerical observations \cite{31}. By characterizing data through shape-related properties such as connected components, loops, and higher-dimensional voids, TDA provides representations that are well suited for knowledge extraction in dynamic and uncertain environments. Recent studies have demonstrated the potential of TDA for improving predictive performance in multivariate time series applications, including occupancy analysis and forecasting tasks in complex systems \cite{32}.

An increasing body of work has explored the integration of TDA into time series analysis and forecasting. Guzel and Kaygun \cite{34} showed that topological feature representations can enhance classification performance even under unbalanced sampling conditions. Karan and Kaygun \cite{35} employed persistent homology in conjunction with time-delay embeddings to extract informative topological features from univariate time series, demonstrating improved robustness to noise and enhanced classification accuracy on physiological datasets. More recent approaches have embedded topological information directly into forecasting architectures. Zeng et al. \cite{36} introduced a topological attention mechanism that incorporates local topological features into attention-based models, achieving improved performance on benchmark datasets. Similarly, Han et al. \cite{37} combined TDA-derived features with deep learning models to enhance short-term solar irradiance forecasting, while Souto \cite{38} linked persistent homology with tail dependence theory to improve volatility forecasting in financial markets.

The efficiency of TDA has been further demonstrated across diverse application domains, including biomedical imaging, financial analytics, and intelligent transportation systems, highlighting its versatility as a data analysis and representation framework \cite{39, 40}. These studies collectively indicate that TDA is particularly well suited for scenarios involving multiscale structure, high noise levels, and complex temporal dynamics—conditions commonly encountered in modern engineering systems.

The core analytical tool underlying most TDA-based approaches is persistent homology, which tracks the evolution of topological features across multiple spatial or temporal scales \cite{41}. Persistent homology distinguishes meaningful structural patterns from noise by identifying features that persist over wide ranges of scales and summarizes them using persistence diagrams or barcodes. Complementary techniques such as the Mapper algorithm provide graph-based representations that facilitate qualitative exploration and dimensionality reduction of complex datasets \cite{42, 43}. Despite these advantages, challenges remain related to computational complexity, parameter sensitivity, and scalability for large-scale or high-dimensional data \cite{44,45}, motivating ongoing research into efficient and interpretable TDA-based representations.

\begin{figure}[H]
	\centering
	\includegraphics[width=0.75\textwidth]{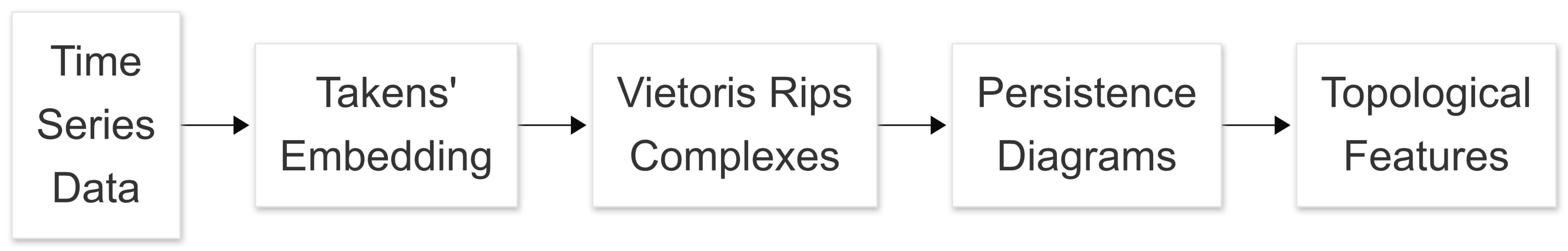}
	\caption{Flowchart of the topological feature extraction}\label{fig:topo}
\end{figure}

Figure~\ref{fig:topo} illustrates a generic topological feature extraction workflow commonly adopted in TDA-based time series analysis \cite{32}. In line with this paradigm, recent studies have increasingly focused on transforming persistence based representations into vectorized features suitable for integration with statistical and machine learning models. Libraries such as Giotto-TDA provide a range of vectorization techniques, including persistence images, Betti curves, and persistence entropy, enabling the incorporation of topological knowledge into data-driven modeling pipelines \cite{46}. These representations have been successfully applied in industrial analytics and smart manufacturing contexts, demonstrating their potential to support knowledge-intensive forecasting and decision-making tasks \cite{47}.

\section{Methodology}
\label{sec2}
This section systematically details the methodological framework and the real production plant data used in this study:

\subsection{Description of Our Methodology}

\begin{figure}[H]
	\centering
	\includegraphics[width=0.7\textwidth]{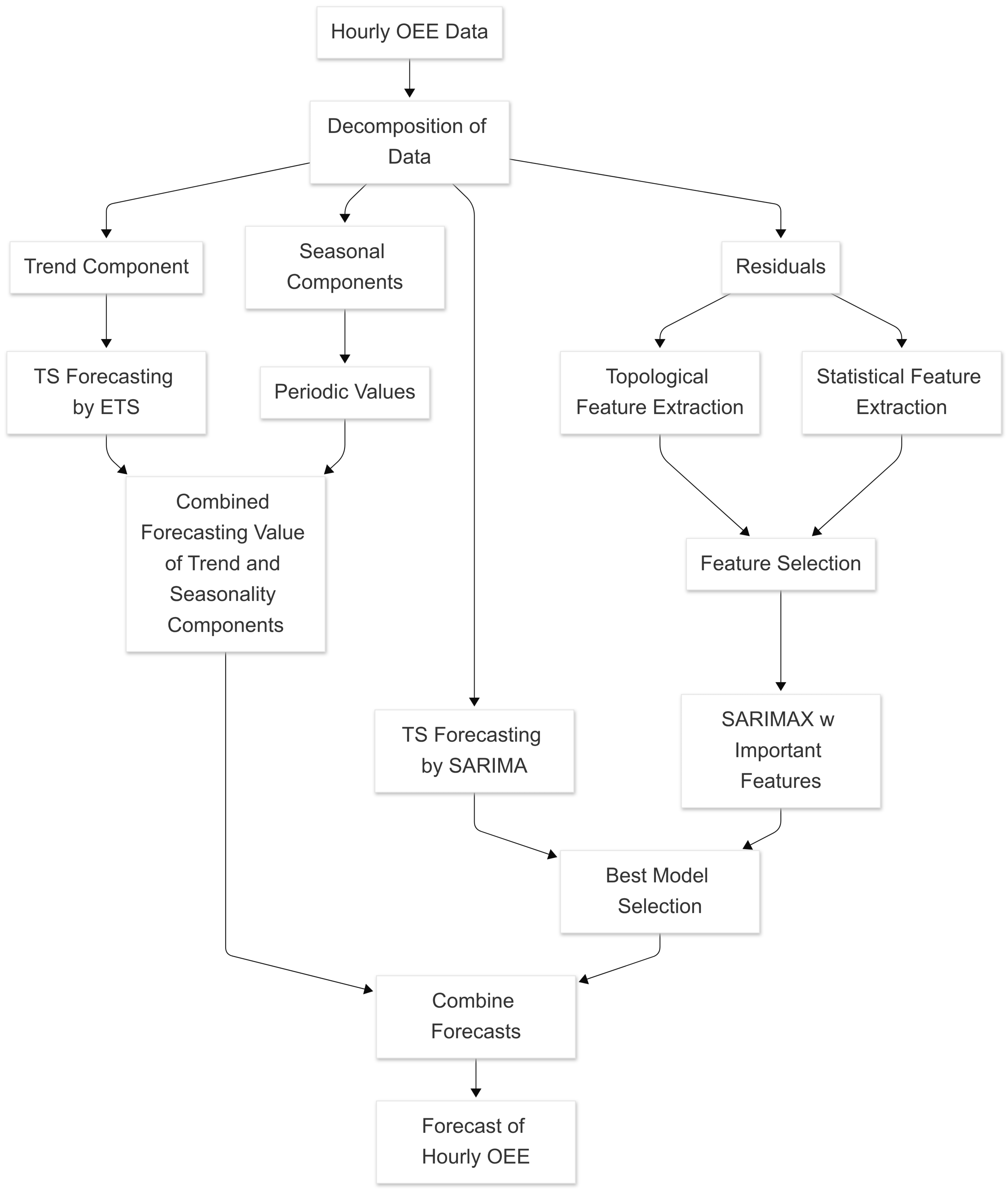}
	\caption{Flowchart of the proposed method}
	\label{fig:algorithm2}
\end{figure}

The proposed methodology is designed to support knowledge-intensive short-term forecasting of highly volatile OEE time series in dynamic manufacturing environments. Rather than treating OEE as a single numerical signal, the framework explicitly represents different sources of temporal behavior by decomposing the original time series into trend, seasonal, and residual components. This decomposition provides a structured view of OEE dynamics and enables component-wise reasoning about long-term evolution, recurring operational patterns, and short-term irregular variations. An overview of the complete methodological pipeline is illustrated in Figure~\ref{fig:algorithm2}.

The trend component captures gradual changes in equipment efficiency associated with long-term operational evolution and is modeled using an Exponential Smoothing State Space (ETS) approach, which provides an interpretable representation of smooth temporal dynamics. Seasonal components reflect periodic production patterns arising from operational schedules, including shift-based, daily, and weekly cycles, and are modeled as recurring temporal structures. The residual component represents short-term operational variability and irregular disturbances that cannot be explained by trend or seasonality alone. Due to its highly volatile nature, this component is modeled using a Seasonal AutoRegressive Integrated Moving Average (S/ARIMA) framework.

To enhance the representation of complex and irregular residual dynamics, the residual forecasting process is augmented with exogenous features derived from both statistical descriptors and topological representations. Statistical features capture conventional temporal properties, while topological features extracted via TDA encode intrinsic structural patterns that are robust to noise and short-term fluctuations. This combination enables a richer and more expressive representation of residual behavior than numerical modeling alone.

The resulting feature space is high-dimensional, making feature selection a critical step for isolating informative representations. To this end, the methodology incorporates a two-stage selection strategy that combines Recursive Feature Elimination (RFE) with Particle Swarm Optimization (PSO). This process systematically filters redundant or uninformative features while preserving those that contribute most strongly to predictive relevance and representational quality.

Forecasting is performed using a Seasonal AutoRegressive Integrated Moving Average with eXogenous variables (SARIMAX) model, in which the selected statistical and topological features serve as external inputs. This formulation enables the model to jointly capture temporal dependencies and structured feature interactions. Forecasts of the trend, seasonal, and residual components are subsequently integrated to produce a comprehensive short-term OEE forecast that reflects multiple levels of temporal behavior.

Model performance is assessed through a systematic evaluation process based on predefined accuracy criteria, including Mean Absolute Error (MAE) and Mean Absolute Percentage Error (MAPE), to identify the most reliable forecasting configuration. In addition, the proposed framework is benchmarked against classical forecasting approaches and recent transformer-based foundation models to contextualize its performance relative to both traditional statistical methods and state-of-the-art deep learning architectures. The application of the proposed methodology across multiple datasets is presented in the following sections.

\subsection{Data}

\begin{figure}[H]
\centering
\begin{subfigure}[b]{0.48\textwidth}
    \centering
    \includegraphics[width=\textwidth]{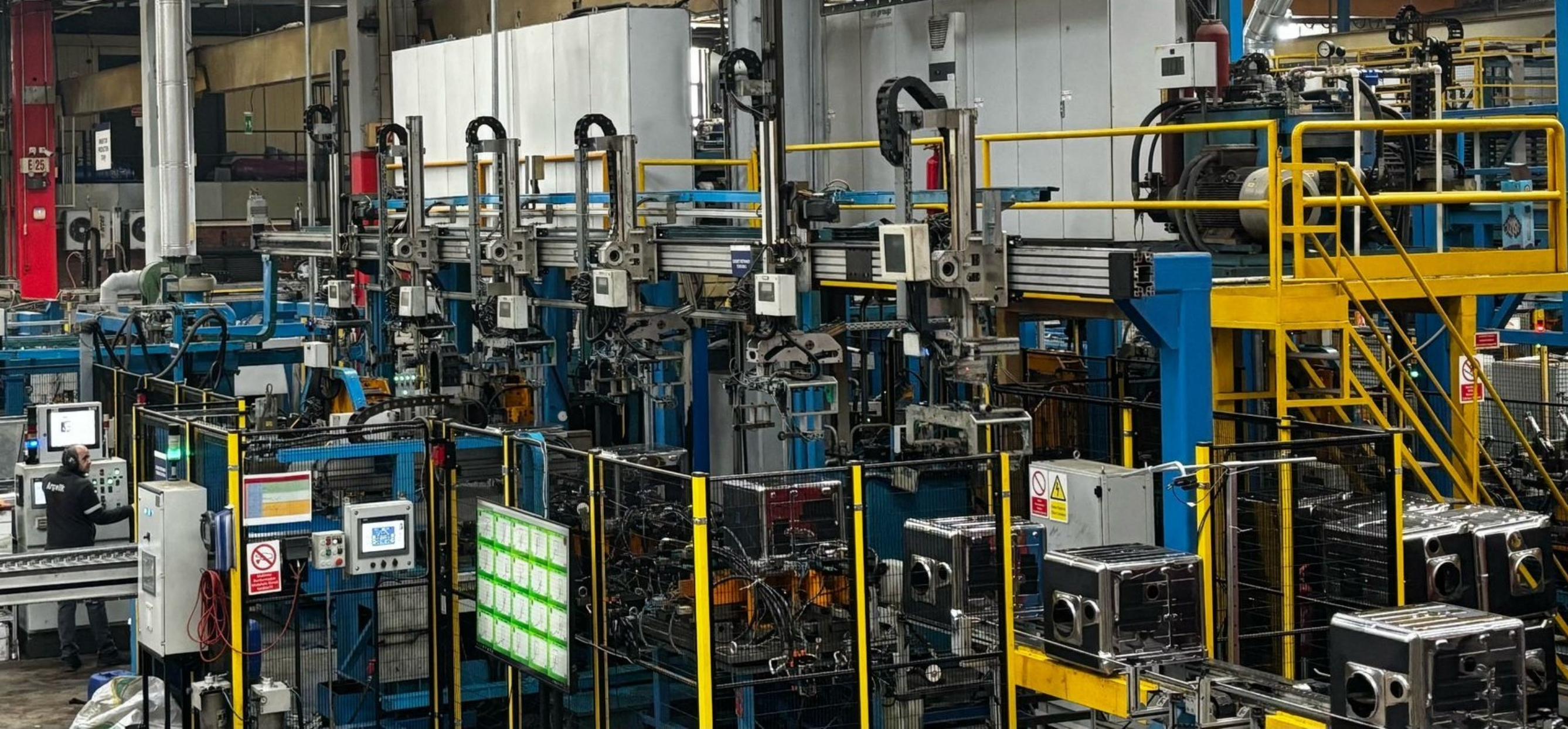}
    \caption{Production equipment GH2}
    \label{fig:gh1}
\end{subfigure}
\hfill
\begin{subfigure}[b]{0.4\textwidth}
    \centering
    \includegraphics[width=\textwidth]{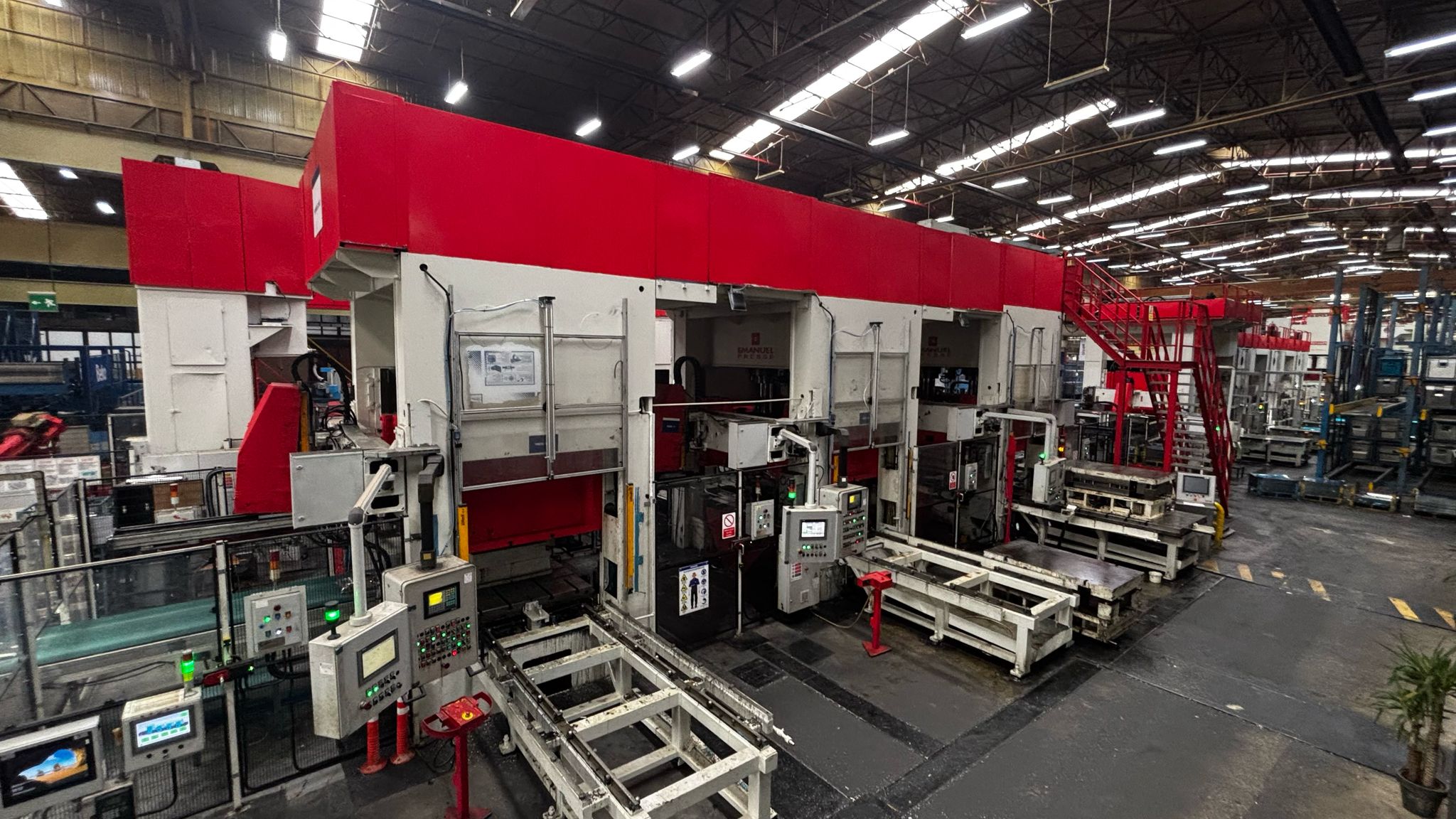}
    \caption{Production equipment H2}
    \label{fig:h1}
\end{subfigure}
\caption{Production equipment used in the study: (a) GH2 system and (b) H2 hydraulic press system.}
\label{fig:equipment}
\end{figure}

The proposed methodology is evaluated using three real-world datasets collected from distinct production equipment, denoted as GH2, H2, and GM2. GH2 and GM2 correspond to advanced, highly complex production systems responsible for manufacturing the inner body group of household appliances, while H2 represents a hydraulic press system comprising four presses used for producing stainless steel body components. Visual illustrations of the production equipment are provided in Figures~\ref{fig:gh1} and~\ref{fig:h1}. The use of heterogeneous equipment enables the assessment of the generality and adaptability of the proposed framework across different manufacturing contexts.

Each dataset consists of hourly OEE measurements with values ranging from 1 to 60, where lower values indicate limited or no operation and higher values correspond to full operational efficiency. The datasets span approximately one month of operation, comprising around 650–680 observations per system, and therefore represent short-term monitoring scenarios typical of real industrial environments. Descriptive statistics for all datasets are summarized in Table~\ref{tab:dataset_summary}, revealing substantial variability across equipment types, with standard deviations exceeding 20 in all cases. This level of variability reflects the dynamic and disturbance-driven nature of manufacturing operations.

\begin{table}[h]
\centering
\resizebox{0.80\textwidth}{!}{%
\begin{tabular}{@{}lccc@{}}
	\toprule
	\textbf{Statistic} & \textbf{GH2 Operation} & \textbf{H2 Operation} & \textbf{GM2 Operation} \\
	\midrule
	Count              & 648                    & 683                   & 672                    \\
	Mean               & 26.58                  & 36.44                 & 30.30                  \\
	Std Dev            & 24.19                  & 21.90                 & 24.86                  \\
	Min                & 1.00                   & 1.00                  & 1.00                   \\
	25th Percentile    & 1.00                   & 16.00                 & 1.00                   \\
	Median (50\%)      & 30.00                  & 46.00                 & 41.00                  \\
	75th Percentile    & 51.00                  & 54.00                 & 54.00                  \\
	Max                & 60.00                  & 60.00                 & 60.00                  \\
	\bottomrule
\end{tabular}}
\caption{Statistical Summary of the Datasets}
\label{tab:dataset_summary}
\end{table}

\begin{figure}[h]
    \centering

    \begin{minipage}[b]{0.45\textwidth}
        \centering
        \includegraphics[width=\textwidth]{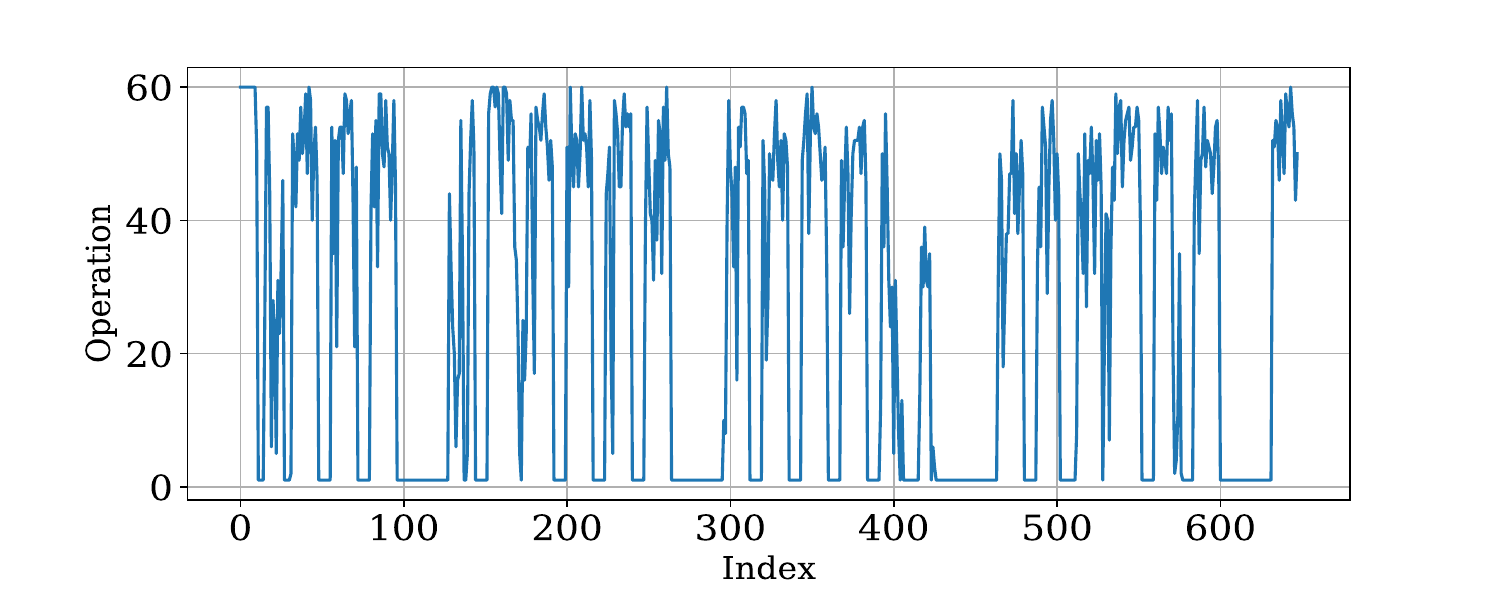}
        \caption*{(a) GH2}
    \end{minipage}
    \hfill
    \begin{minipage}[b]{0.45\textwidth}
        \centering
        \includegraphics[width=\textwidth]{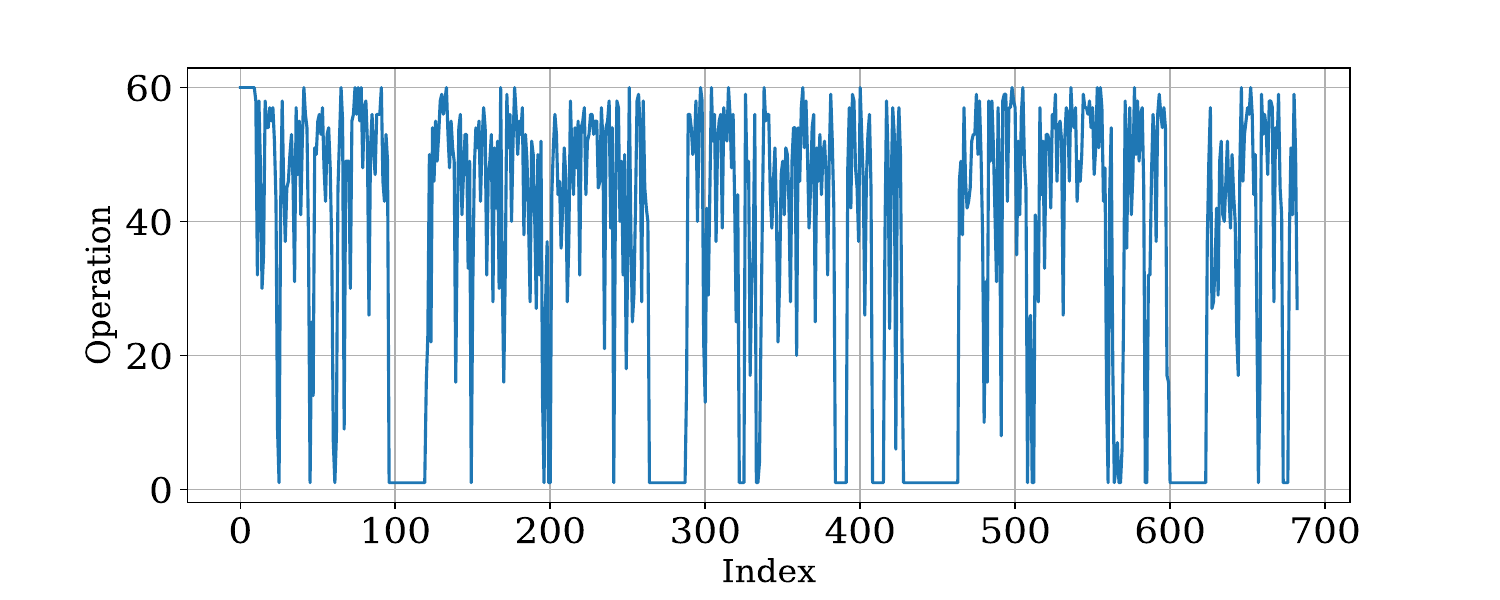}
        \caption*{(b) H2}
    \end{minipage}

    \vspace{1em}

    \begin{minipage}[b]{0.45\textwidth}
        \centering
        \includegraphics[width=\textwidth]{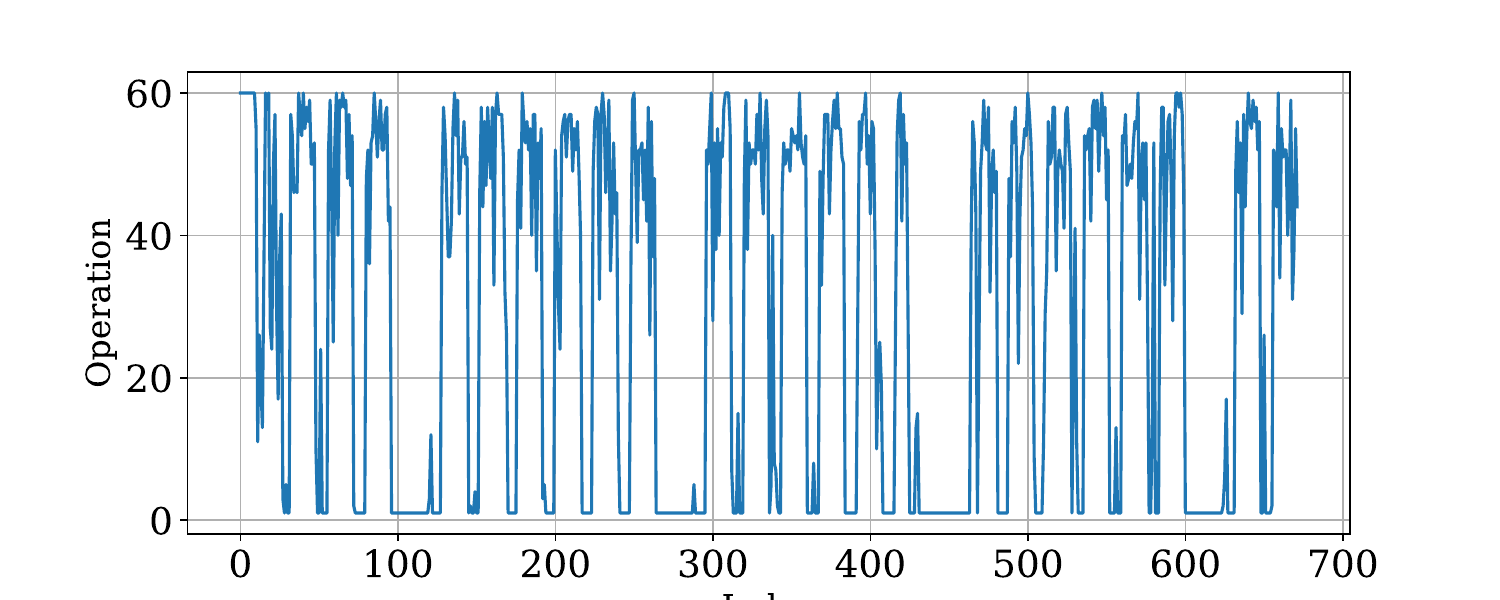}
        \caption*{(c) GM2}
    \end{minipage}

    \caption{Time series plots of hourly OEE values (1–60) for GH2, H2 and GM2 datasets.}
    \label{fig:ts_all}
\end{figure}

Figure~\ref{fig:ts_all} presents the time series plots of the OEE data for GH2, H2, and GM2. The series exhibit pronounced fluctuations, abrupt changes, and intermittent periods of low utilization. 

\begin{figure}[H]
	\centering
	\includegraphics[width=0.35\textwidth]{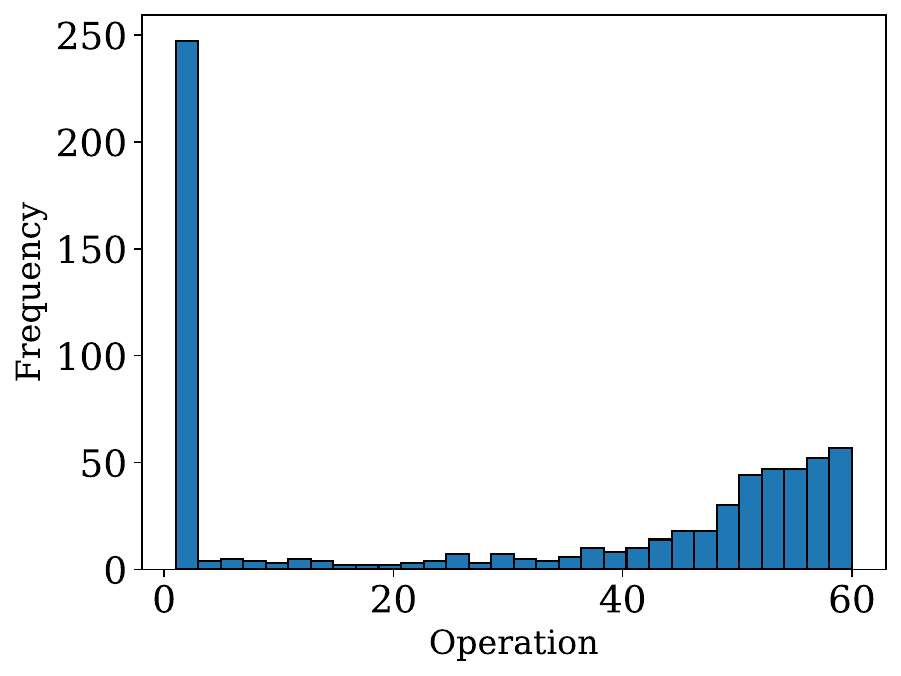}
	\caption{Histogram of OEE values of equipment GM2}\label{fig:hist_h1}
\end{figure}

In addition, the distributions of OEE values deviate from normality, as illustrated by the histogram shown in Figure~\ref{fig:hist_h1} for the GM2 dataset, highlighting the presence of skewness and heavy-tailed behavior. Such characteristics pose significant challenges for conventional forecasting approaches based on Gaussian assumptions.

\begin{figure}[H]
	\centering
	
	\begin{minipage}[b]{0.3\textwidth}
		\centering
		\includegraphics[width=\textwidth]{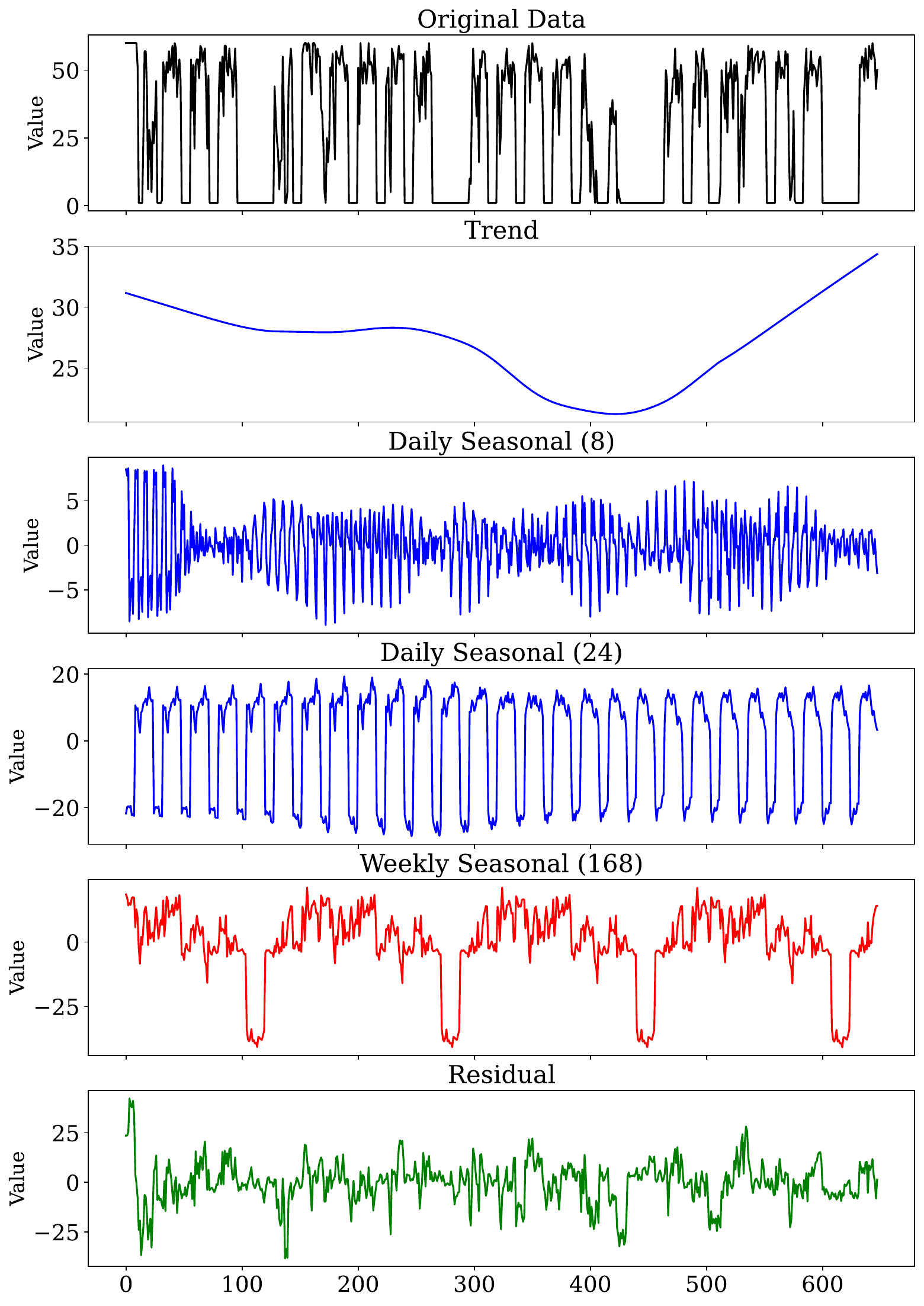}
		\caption*{(a) GH2}
	\end{minipage}
	\hfill
	\begin{minipage}[b]{0.3\textwidth}
		\centering
		\includegraphics[width=\textwidth]{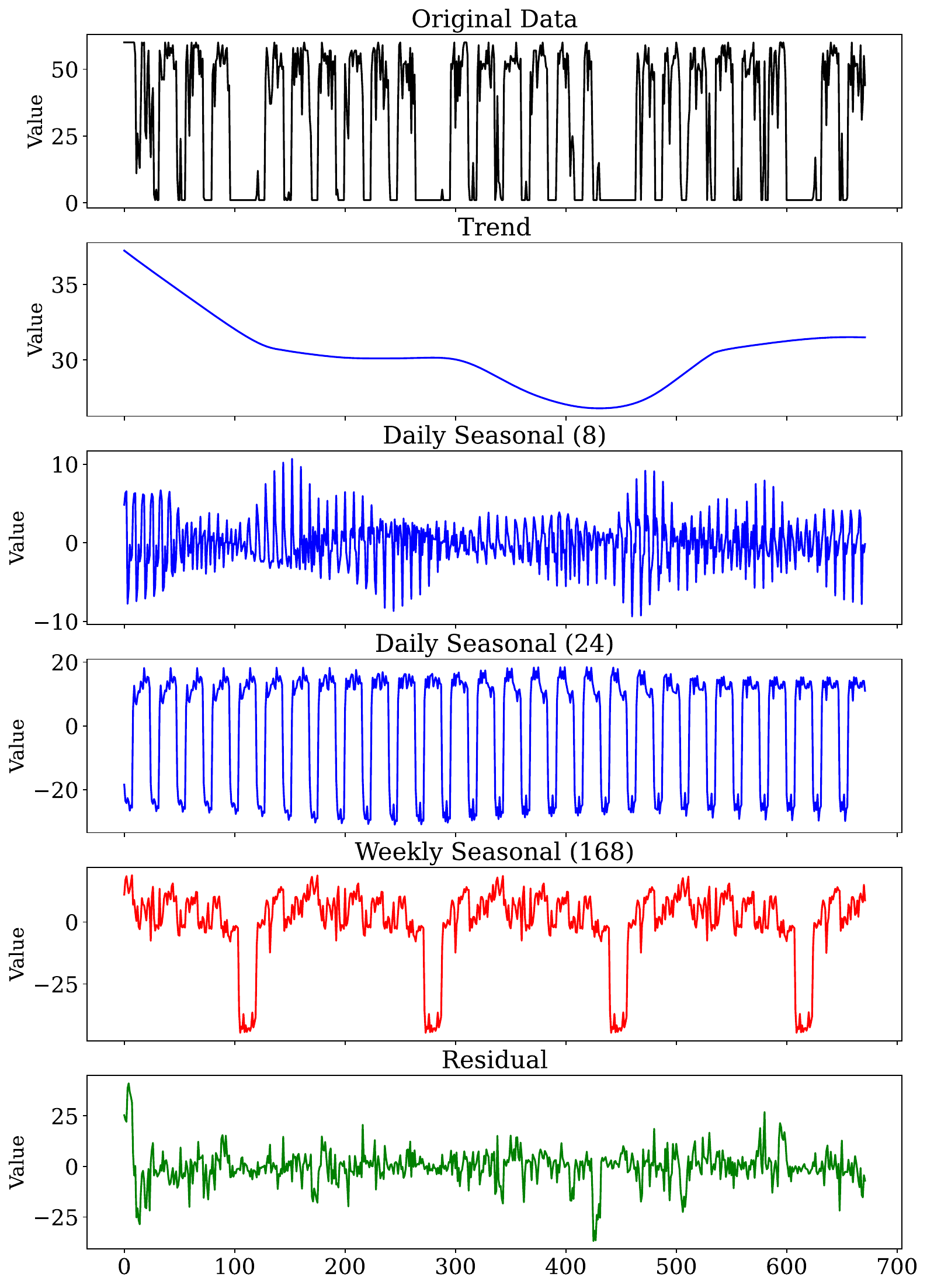}
		\caption*{(b) GM2}
	\end{minipage}
	\hfill
	\begin{minipage}[b]{0.3\textwidth}
		\centering
		\includegraphics[width=\textwidth]{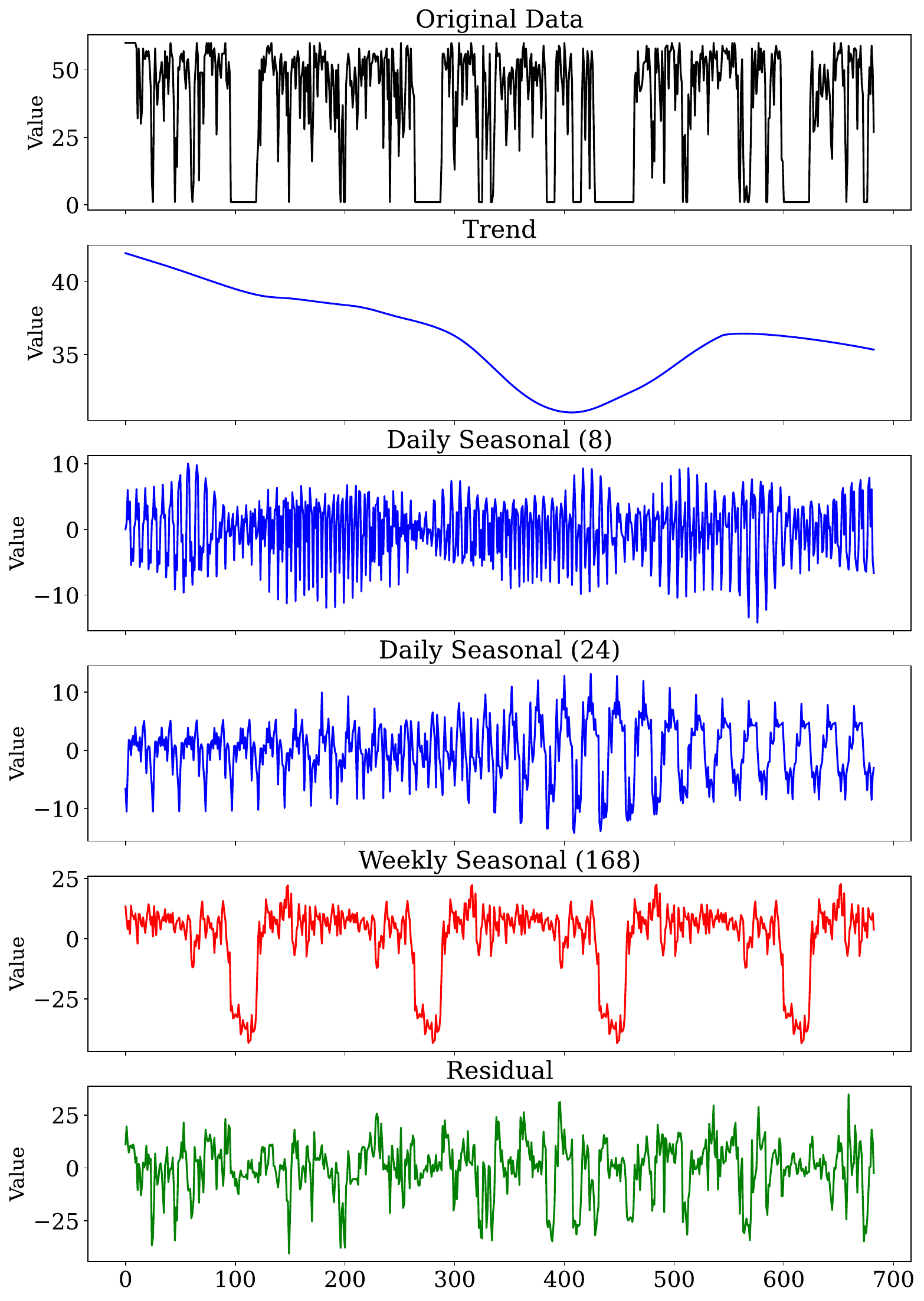}
		\caption*{(c) H2}
	\end{minipage}
	
	\caption{Time series decomposition of GH2, GM2, and H2 datasets.}
	\label{fig:decomposition}
\end{figure}

From a domain perspective, the OEE time series are characterized by multiple overlapping seasonal patterns associated with production schedules. In particular, shift-based (8-hour), daily (24-hour), and weekly (168-hour) periodicities are expected due to operational routines and workforce organization. These multi-seasonal structures are evident in the data and motivate the use of decomposition-based representations to separate long-term trends, recurring operational cycles, and short-term irregular behavior. Representative decomposition results are illustrated in Figure~\ref{fig:decomposition}.

Overall, the datasets exhibit a combination of short temporal horizons, high volatility, non-normal distributions, and complex seasonal structure. These properties make them well suited for evaluating forecasting methodologies that emphasize expressive temporal representations and robust knowledge extraction. In the following sections, statistical and topological feature extraction techniques are employed to derive exogenous variables that capture these characteristics and support the proposed SARIMAX-based forecasting framework.


\section{Feature Extraction and Selection for Data Knowledge and Forecasting}

Forecasting highly volatile and short-term OEE time series requires representations that extend beyond raw numerical observations. While time series values provide instantaneous measurements of equipment efficiency, they do not explicitly encode structural properties such as variability patterns, temporal dependencies, or multiscale dynamics that are essential for knowledge-based reasoning and reliable forecasting.

Feature-based representations provide a principled mechanism for transforming raw time series data into structured descriptors that summarize relevant temporal characteristics. By extracting informative features, complex temporal behavior can be represented in a form that is more amenable to interpretation, comparison, and integration into forecasting models. In this study, feature extraction is viewed not merely as a preprocessing step, but as a process of knowledge representation that enables reasoning about operational behavior under uncertainty.

To capture complementary aspects of OEE dynamics, two distinct classes of features are considered. Statistical features describe local numerical properties and conventional temporal characteristics, while topological features encode global and multiscale structural information that remains robust under noise and short-term fluctuations. Given the high dimensionality of the resulting feature space, a dedicated feature selection strategy is employed to filter redundant or uninformative representations and retain features that contribute most strongly to predictive relevance and knowledge expressiveness.

The following subsections describe the statistical and topological feature representations in detail, followed by the feature selection methodology used to support robust and interpretable short-term OEE forecasting.
\subsection{Statistical Feature Extraction}

Statistical feature extraction is employed to provide baseline numerical representations of short-term residual OEE dynamics. Using the \texttt{tsfresh} library, a comprehensive set of time-domain and frequency-domain descriptors is extracted from the residual component of the OEE time series \cite{48}. These features capture local statistical properties and conventional temporal characteristics that are commonly used in data-driven forecasting models.

The extracted features can be grouped into several categories according to the type of temporal information they represent.

\textit{Descriptive and deviation metrics} summarize the central tendency, dispersion, and magnitude of variations within the residual series, including measures such as mean, variance, skewness, kurtosis, absolute energy, and first-order difference--based change indicators. These features provide a compact numerical description of short-term variability.

\textit{Frequency-domain features} are derived via Fourier analysis and characterize periodic behavior and dominant frequency components present in the residual signal. This group includes Fourier coefficients and aggregated statistics computed over the frequency spectrum, which help identify recurrent patterns and oscillatory behavior \cite{49}. 

\textit{Autocorrelation and temporal dependency measures} quantify persistence and repeating structures in the residual series by aggregating autocorrelation-based statistics over multiple lags. These descriptors capture linear temporal dependencies and are commonly used in classical time series modeling.

\textit{Entropy and complexity measures}, including entropy-based descriptors such as sample entropy and approximate entropy, are extracted to assess the degree of irregularity and unpredictability in the residual dynamics \cite{49,50}. Such measures provide insight into signal complexity but remain sensitive to noise and local fluctuations.

Finally, \textit{trend and change quantile features} describe systematic directional changes and distributional shifts over time, offering additional numerical summaries of residual behavior. While these statistical features provide valuable local descriptors, they primarily encode pointwise and aggregated information and do not explicitly represent the global or multiscale structure of the underlying temporal dynamics. This limitation motivates the incorporation of topological representations, described in the following subsection.

\subsection{Topological Feature Extraction}

The proposed methodology leverages TDA to extract informative and noise-robust feature representations from time series data, thereby enhancing forecasting accuracy and supporting knowledge-driven analytical applications \cite{51, 52}. This section describes the overall extraction pipeline and the mathematical foundations of the employed topological descriptors.

Initially, the raw time series is converted into a numerical array and segmented into overlapping windows using a sliding window transformation with a window size of 24 and a stride of 1 \cite{53}. This operation preserves local temporal dynamics by isolating short-term behavioral patterns within each window. Each window is subsequently embedded into a higher-dimensional phase space via Takens embedding \cite{54}, using a fixed time delay of 8 and an embedding dimension of 3. The time delay parameter, denoted by $\tau$, is selected to ensure sufficient independence between successive embedding coordinates, thereby unfolding the underlying system dynamics without redundancy \cite{30}.

The time delay $\tau$ is commonly determined using either the autocorrelation function, by selecting the first zero crossing or the point where the autocorrelation decays to $1/e$ of its initial value, or via the average mutual information method by identifying the first local minimum \cite{53}. Similarly, the embedding dimension $m$ is typically estimated using the False Nearest Neighbors algorithm or Cao’s method, which detect the minimal dimension required to avoid projection-induced artifacts \cite{55}. In this study, representative embedding parameters are selected to ensure consistency across windows \cite{32}.

\begin{figure}[H]
	\centering
	\includegraphics[width=1.0\textwidth]{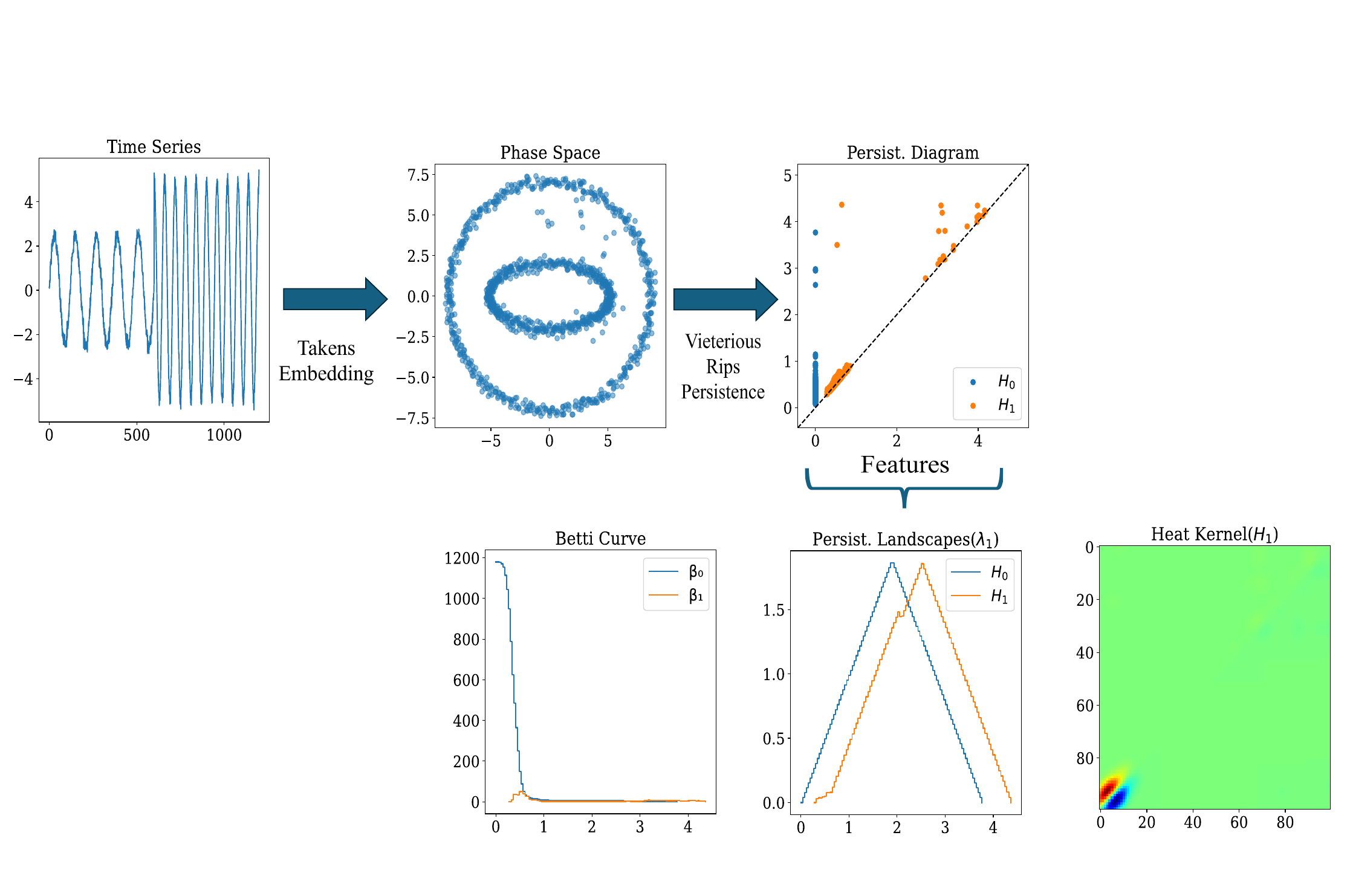}
	\caption{The workflow of TDA feature extraction}
	\label{fig:giotto}
\end{figure}

Following phase space reconstruction, the Vietoris Rips persistence method is applied to each embedded point cloud \cite{56, 53}. As the filtration scale varies, this method tracks the emergence and disappearance of topological features such as connected components and loops across homology dimensions $H_0$ and $H_1$ \cite{30}. The resulting persistence diagrams encode the multiscale topological structure of the time series. To ensure comparability across samples, the diagrams are appropriately scaled. The standard TDA feature extraction workflow is illustrated in Figure~\ref{fig:giotto} \cite{55, 57, 58}.

A diverse set of TDA-based feature extractors is then applied to the scaled persistence diagrams to obtain structured, vectorized representations:

\begin{itemize}

\item \textbf{Persistence Entropy:}  
Persistence entropy quantifies the complexity of a persistence diagram by computing the Shannon entropy of feature lifetimes \cite{53, 58, 59, 60}. 

Let \( D \) be a persistence diagram consisting of a set of persistence pairs \( p_i = (b_i, d_i) \), where \( b_i \) is the birth time and \( d_i \) is the death time of a topological characteristic. The lifetime of each feature is
\begin{equation}
	l_i = d_i - b_i.
\end{equation}
The normalized lifetime distribution is defined as
\begin{equation}
	p_i = \frac{l_i}{\sum_j l_j},
\end{equation}
and the persistence entropy is given by
\begin{equation}
	H = -\sum_i p_i \log p_i.
\end{equation}

\item \textbf{Amplitude Metrics:}  
Amplitude descriptors summarize the overall persistence content of a \emph{single} persistence diagram by applying norm-like functionals to feature lifetimes \cite{61}. These measures capture the magnitude of topological activity without comparing diagrams:
\begin{itemize}
	\item \textbf{Bottleneck Amplitude:}
	\begin{equation}
		\mathrm{Amp}_B(D) = \max_{(b_i,d_i)\in D}(d_i-b_i),
	\end{equation}
	representing the most dominant topological feature \cite{62}.
	\item \textbf{Wasserstein Amplitude:}
	\begin{equation}
		\mathrm{Amp}_{W,p}(D) =
		\left(
		\sum_{(b_i,d_i)\in D}(d_i-b_i)^p
		\right)^{1/p},
	\end{equation}
	which aggregates persistence across all features \cite{55}.
\end{itemize}

\item \textbf{Betti Curves and Persistence Landscapes:}  
\begin{itemize}
	\item \textbf{Betti Curves:}  
	Betti curves track the number of topological features as a function of the filtration parameter $t$:
	\begin{equation}
		\mathcal{B}_n(t)=\sum_{(b_i,d_i)\in D_n}\mathbf{1}_{[b_i,d_i)}(t),
	\end{equation}
	where $\beta_0$, $\beta_1$, and $\beta_2$ correspond to connected components, loops, and voids, respectively \cite{61, 63}.
	\item \textbf{Persistence Landscapes:}  
	Persistence landscapes provide stable functional summaries of persistence diagrams:
	\begin{equation}
		\lambda_k(t)=\sup_{\substack{(b_i,d_i)\in D \\ k\text{-th largest}}}
		\max(0,\min(t-b_i,d_i-t)),
	\end{equation}
	with the associated $L^p$ norm \cite{64}
	\begin{equation}
		L_p=\left(\int_{-\infty}^{\infty}\sum_k|\lambda_k(t)|^pdt\right)^{1/p}.
	\end{equation}
\end{itemize}

\item \textbf{Silhouette and Heat Kernel Representations:}  
\begin{itemize}
	\item \textbf{Silhouette Representation:}  
	The silhouette provides a weighted average of persistence landscapes:
	\begin{equation}
		S_\alpha(t)=
		\frac{\sum_{(b_i,d_i)\in D}(d_i-b_i)^\alpha\phi_i(t)}
		{\sum_{(b_i,d_i)\in D}(d_i-b_i)^\alpha},
	\end{equation}
	where $\phi_i(t)=\max(0,\min(t-b_i,d_i-t))$ and $\alpha\ge0$ controls feature weighting \cite{53, 63}.
	\item \textbf{Heat Kernel Representation:}  
	This representation smooths persistence diagrams using a diffusion process with a diagonal reflection to suppress noise:
	\begin{equation}
		\Phi_\sigma(z)=
		\frac{1}{4\pi\sigma}
		\sum_{(b_i,d_i)\in D}
		\left(
		e^{-\frac{\|z-p_i\|^2}{4\sigma}}-
		e^{-\frac{\|z-\bar p_i\|^2}{4\sigma}}
		\right),
	\end{equation}
	where $\bar p_i=(d_i,b_i)$ \cite{57}.
\end{itemize}

\end{itemize}

Each extractor produces a vectorized feature representation that is flattened and organized into a tabular format. In addition, summary statistics of feature lifetimes, including total, mean, variance, and extrema, are computed for each homology dimension \cite{52, 58}. All extracted features are concatenated into a comprehensive feature set, which is subsequently refined through variance-based and correlation-based feature selection. The resulting representation captures both the temporal dynamics and the underlying topological structure of the time series, supporting accurate and robust forecasting \cite{51, 57}.

\subsection{Feature Selection}

The initial feature extraction stage yields more than 450 statistical and topological features. While this rich representation captures diverse temporal and structural characteristics of OEE dynamics, such high dimensionality introduces the curse of dimensionality, increases redundancy, and can degrade short-term forecasting performance. To address these challenges, a multi-stage feature selection framework is employed to reduce dimensionality while preserving the most informative predictors for forecasting.

The selection process begins with low-variance filtering, where features with variance below 0.01 are removed. This step eliminates non-informative descriptors that contribute little explanatory power as exogenous variables in SARIMAX models. Subsequently, multicollinearity is addressed by identifying and removing highly dependent feature columns, thereby improving numerical stability and interpretability.

To further refine the feature set, a correlation-aware selection mechanism is applied. Rather than discarding correlated features arbitrarily, feature importance is evaluated using a Random Forest regressor trained on the forecasting task. Among strongly correlated features, those with lower predictive contribution are eliminated. After these filtering stages, the feature space is reduced to approximately 150 features for both statistical and topological representations. Forecasting experiments are conducted at this stage, and if dimensionality remains excessive, additional model-driven selection methods are applied.

\subsubsection{Feature Selection via SARIMAX Model Performance}

The final feature selection stage is explicitly aligned with forecasting objectives and is based on statistical significance and model parsimony, quantified using the Bayesian Information Criterion (BIC). The procedure consists of two steps: recursive elimination based on SARIMAX p-values, followed by Particle Swarm Optimization (PSO) for BIC minimization.

\textit{First step: Recursive elimination via p-values.}  
A SARIMAX model is fitted using the available exogenous features, and p-values are computed for each predictor. Features with p-values exceeding 0.05 are considered statistically insignificant and are recursively removed. At each iteration, the model is refitted and significance is re-evaluated until all retained features satisfy the threshold or a minimum feature count is reached. This ensures that only statistically meaningful predictors remain in the forecasting model.

\begin{figure}[h!]
\centering
\resizebox{\textwidth}{!}{%
\begin{tikzpicture}[
  font=\small,
  >=Latex,
  node distance=18mm and 16mm,
  block/.style={
    draw,
    rounded corners,
    align=center,
    minimum width=32mm,
    minimum height=12mm
  },
  tallblock/.style={
    draw,
    rounded corners,
    align=center,
    minimum width=30mm,
    minimum height=30mm
  },
  decision/.style={
    draw,
    diamond,
    aspect=2.2,
    align=center,
    inner sep=2pt
  },
  line/.style={-Latex, thick}
]

\node[tallblock] (A) {Residual\\Component\\of\\Decomposition};

\node[block, right=of A, yshift=10mm] (B) {Topological\\Features};
\node[block, right=of A, yshift=-10mm] (C) {Statistical\\Features};

\node[tallblock, right=of A, xshift=45mm] (D)
{Initial Feature\\Selection\\[2pt]
Low Variance \&\\
Multicollinearity\\
Removal};

\node[decision, right=of D, xshift=5mm] (E)
{Apply RFE \&\\ PSO?};

\node[tallblock, right=of E, yshift=0mm, xshift=100mm] (H)
{Final\\Exogenous\\Variables};

\node[block, right=of E, yshift=-18mm, xshift=5mm] (F)
{RFE\\(SARIMAX\\p-value)};
\node[block, right=of F] (G)
{PSO\\(Minimize BIC)};

\draw[line] (A.east) -- (B.west);
\draw[line] (A.east) -- (C.west);

\draw[line] (B.east) -- (D.west);
\draw[line] (C.east) -- (D.west);

\draw[line] (D.east) -- (E.west);

\draw[line] (E.east) -- node[above]{no} (H.west);
\draw[line] (E.south east) -- node[below]{yes} (F.west);
\draw[line] (F.east) -- (G.west);
\draw[line] (G.east) -- (H.west);

\end{tikzpicture}%
}
\caption{Flowchart of the feature selection methods}
\label{fig:fsel}
\end{figure}
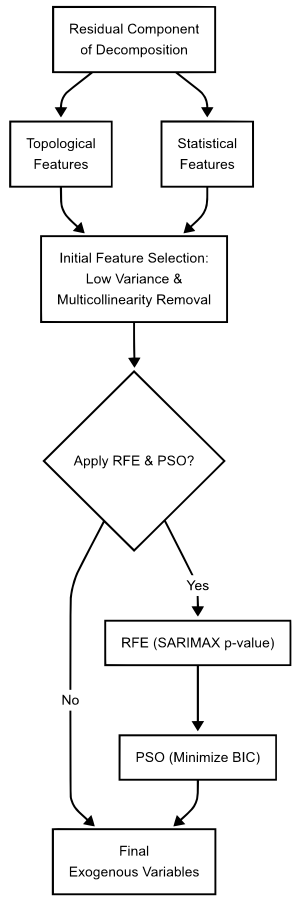

\textit{Second step: Feature elimination via PSO using BIC.}  
If the feature space remains large after recursive elimination, PSO is employed for further refinement. Each particle represents a candidate feature subset encoded as a binary vector, and the BIC of a SARIMAX model fitted on that subset serves as the objective function. The PSO configuration uses a swarm size of 40 particles, a maximum of 300 iterations, an inertia weight of $\omega = 0.7$, and cognitive and social parameters of $\phi_p = 1.4$ and $\phi_g = 1.8$. The optimization is executed fifteen times to ensure robustness. A flowchart of the selection process is shown in Figure~\ref{fig:fsel}.

This hybrid selection framework substantially reduces dimensionality while improving forecasting stability and interpretability. Beyond numerical reductions, a consistent qualitative pattern emerges across all datasets (GH2, H2, and GM2): Heat Kernel (HK) features constitute the majority of retained topological predictors after both elimination stages.

The persistence of HK features through aggressive filtering reflects their ability to encode stable, noise-robust, and multi-scale topological information via diffusion-based representations. Unlike other topological summaries, HK features consistently provide non-redundant and statistically significant contributions to SARIMAX forecasting performance. Their dominance indicates that Heat Kernel representations capture structurally meaningful temporal information that is particularly well-suited for modeling short-term, volatile OEE dynamics.

\begin{table}[h]
\centering
\resizebox{0.80\textwidth}{!}{%
\begin{tabular}{@{}lccc@{}}
	\toprule
	 & \textbf{GH2 Dataset} & \textbf{H2 Dataset} & \textbf{GM2 Dataset} \\
	\midrule
	\textbf{Statistical Features} & 267 & 268 & 264 \\
	\quad RFE Elimination         & 57  & 48  & 37  \\
	\quad PSO Elimination         & 8   & 11  & 14  \\
	\textbf{Topological Features} & 144 & 143 & 131 \\
	\quad RFE Elimination         & 19  & 55  & 52  \\
	\quad PSO Elimination         & 3   & 5   & 6   \\
	\bottomrule
\end{tabular}}
\caption{Feature Summary for GH2, H2, and GM2 Datasets}
\label{tab:features}
\end{table}


\section{Forecasting Models}
\label{sec:forecasting_models}
Forecasting models play a critical role in this study by operationalizing the extracted statistical and topological representations into actionable predictions of short-term OEE behavior. Consequently, this study adopts a comparative forecasting framework that includes classical statistical models, advanced seasonal models, and recent transformer-based foundation models. This benchmark-oriented design enables a systematic assessment of the proposed SARIMAX-based framework against established and state-of-the-art approaches, with particular emphasis on short-term accuracy, robustness under volatility, and suitability for knowledge-intensive decision support in manufacturing systems.

\subsection{Benchmark Model Set}

To evaluate the efficiency of the proposed forecasting framework, a comprehensive benchmark model set is defined. As a baseline, a SARIMA model is applied directly to the GH2, H2, and GM2 datasets without time-series decomposition. In addition, both classical statistical forecasting models and recent transformer-based foundation models are considered to provide a rigorous comparative evaluation.

The classical benchmark models include Exponential Smoothing (ETS) and TBATS, which are widely used in industrial forecasting applications. The advanced benchmark group consists of recent foundation models for time series forecasting, namely CHRONOS, TimesFM, and Lag-Llama.

\subsubsection{Exponential Smoothing (ETS)}

Exponential Smoothing (ETS) models represent a family of forecasting methods structured around three components: \emph{Error}, \emph{Trend}, and \emph{Seasonality}. The error component may be additive or multiplicative, the trend component may be absent or evolving, and the seasonal component may follow additive or multiplicative dynamics \cite{14, 65}. Owing to their adaptive nature and reliance on local temporal information, ETS models are particularly effective for short-term forecasting and demonstrate robustness in noisy operational environments.

\subsubsection{TBATS Model}

The TBATS model is designed to handle time series exhibiting multiple or non-integer seasonal patterns. The acronym TBATS denotes its key components: trigonometric seasonality representation, Box–Cox transformation for variance stabilization, ARMA error structure, trend modeling, and complex seasonal components \cite{66}. This structure enables TBATS to capture overlapping seasonalities commonly observed in manufacturing systems, such as daily and weekly operational cycles.

\subsubsection{Recent Advancements in Time-Series Forecasting Models}

Recent developments in time-series forecasting increasingly leverage large-scale transformer architectures originally developed for natural language processing to model long-range temporal dependencies. In this study, the following foundation models are considered:

\begin{itemize}
    \item \textbf{Chronos:} Chronos transforms continuous time series into discrete token sequences through scaling and quantization and employs transformer-based language models (T5 variants) trained with a cross-entropy loss. Pretrained on diverse public and synthetic datasets, Chronos demonstrates strong zero-shot performance and adapts efficiently via fine-tuning \cite{67}.
    
    \item \textbf{Lag-Llama:} Lag-Llama is a decoder-only transformer architecture designed for univariate probabilistic time-series forecasting. By incorporating lagged covariates, it effectively captures long-range dependencies and demonstrates robust generalization, particularly when fine-tuned on relatively small datasets \cite{68}.
    
    \item \textbf{TimesFM:} TimesFM is a decoder-only forecasting model developed for fast, out-of-the-box prediction with minimal preprocessing. Its compact architecture enables efficient inference, making it suitable for real-time forecasting applications \cite{69}.
\end{itemize}

\subsection{Proposed Model}

The proposed forecasting framework integrates SARIMAX modeling with statistical and topological feature representations. Each OEE time series is decomposed into seasonal, trend, and residual components. The seasonal and trend components are forecast using classical time-series models, while the residual component—capturing short-term volatility and irregular dynamics—is modeled using SARIMAX with exogenous variables.

\begin{figure}[H]
	\centering
	\includegraphics[width=0.85\textwidth]{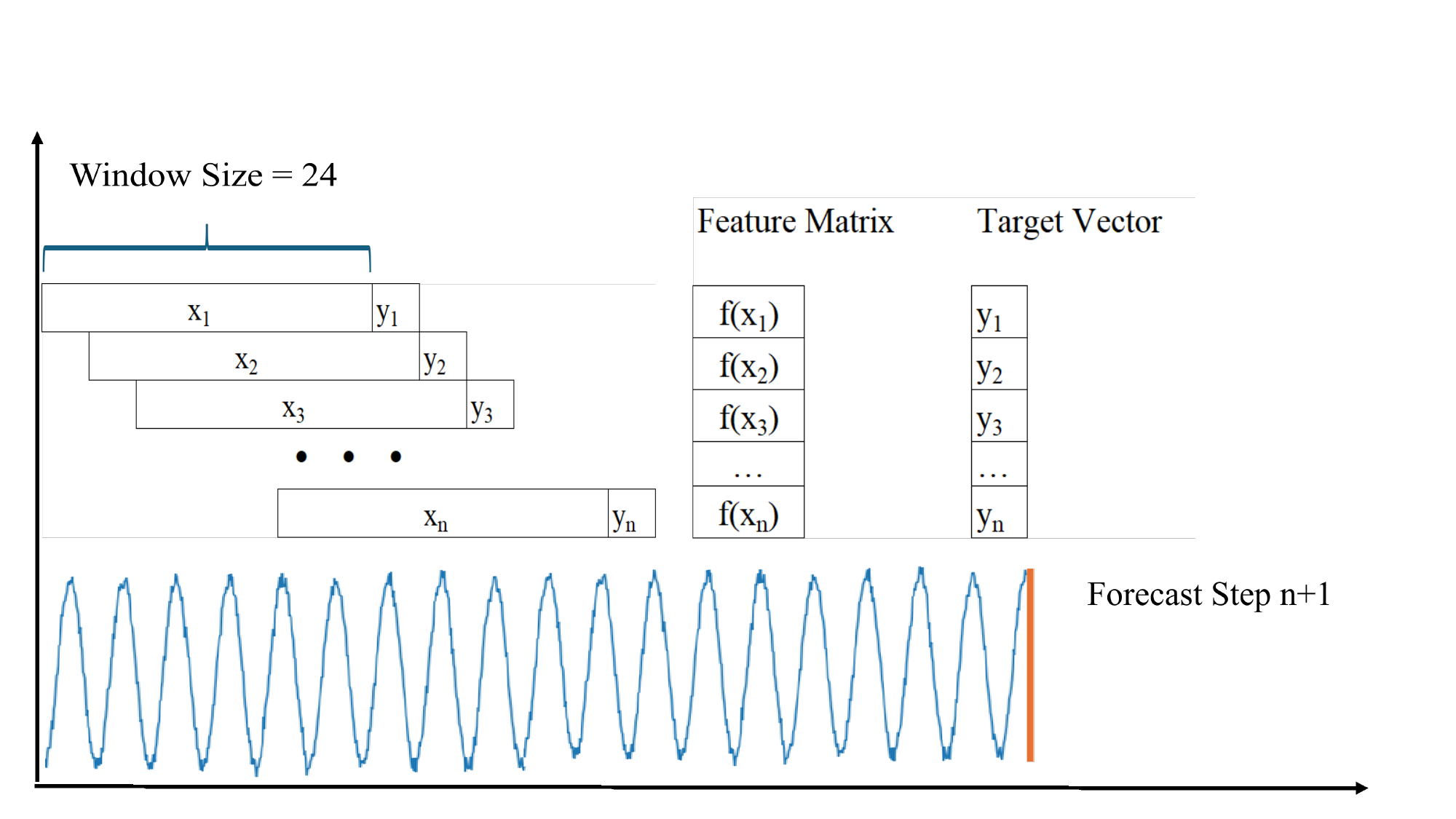}
	\caption{The forecasting strategy for the proposed model}
	\label{fig:forecasting}
\end{figure}

Exogenous features are extracted from sliding windows of length 24. To prevent data leakage, each forecast is generated using only features computed from the preceding window, ensuring strict temporal causality \cite{30}. The overall forecasting workflow is illustrated in Figure~\ref{fig:forecasting}.

A rigorous two-stage feature elimination process is applied to retain only the most informative statistical and topological predictors. This hybrid representation enables the model to capture both long-term temporal dynamics and short-term irregularities inherent in volatile OEE data, thereby enhancing forecasting robustness.

\subsubsection{SARIMAX Modelling}

SARIMAX models serve as the core forecasting mechanism for the residual component. The selection of autoregressive and moving-average orders is guided by analysis of the autocorrelation function (ACF) and partial autocorrelation function (PACF) of the residual series obtained after decomposition.

\begin{itemize}
	\item For the GH2 dataset, the selected model is:
	\[
		\text{SARIMAX}(4,0,0)(1,0,1)_8.
	\]
	\item For the GM2 dataset, the selected model is:
	\[
		\text{SARIMAX}(2,0,0)(2,0,1)_8.
	\]
	\item For the H2 dataset, the selected model is:
	\[
		\text{SARIMAX}(1,0,0)(1,0,1)_8.
	\]
\end{itemize}

For all datasets, standard SARIMAX model assumptions are verified and satisfied for the selected configurations.

\section{Results}

The performance of the benchmark models is presented in Table~\ref{tab:results2}, which reports the MAE, MAPE and computational cost (measured as operating time on a standard Intel i7 processor) for each forecasting method. Model evaluation is based primarily on the MAE and MAPE metrics.

\begin{table}[h!]
\centering
\resizebox{\textwidth}{!}{%
\begin{tabular}{lccccccccc}
\toprule
\textbf{Dataset:} & \multicolumn{3}{c}{\textbf{GH2}} & \multicolumn{3}{c}{\textbf{H2}} & \multicolumn{3}{c}{\textbf{GM2}} \\
\cmidrule(lr){2-4} \cmidrule(lr){5-7} \cmidrule(lr){8-10}
\textbf{Model} & \textbf{MAE} & \textbf{MAPE} & \textbf{Cost} & \textbf{MAE} & \textbf{MAPE} & \textbf{Cost} & \textbf{MAE} & \textbf{MAPE} & \textbf{Cost} \\
\midrule
CHRONOS    &  4.56 & 0.10 & 10 & 12.50 & 0.32 & 9  & 12.10 & 0.34 & 9  \\
Lag-Llama  &  4.76 & 0.10 & 8  & 14.32 & 0.31 & 9  & 11.84 & 0.33 & 9  \\
TimesFM    &  7.37 & 0.14 & 7  & 23.64 & 0.48 & 7  & 14.28 & 0.40 & 10 \\
\midrule
ETS        & 14.49 & 0.28 & 5  &  8.00 & 0.19 & 6  &  9.42 & 0.17 & 6  \\
TBATS      & 18.43 & 0.36 & 38 &  9.80 & 0.22 & 40 &  8.87 & 0.17 & 42 \\
SARIMA     &  3.81 & 0.08 & 5  & 14.11 & 0.29 & 33 & 11.26 & 0.31 & 4  \\
\bottomrule
\end{tabular}%
}
\caption{Comparison of forecasting models without decomposition for GH2, H2, and GM2 datasets}
\label{tab:results2}
\end{table}

In this benchmark setting, no decomposition was applied to the OEE datasets. Across all datasets, modern transformer-based forecasting models (Chronos, Lag-Llama and TimesFM) did not outperform the traditional forecasting approaches. For each dataset, at least one classical or statistical model (ETS, TBATS or SARIMA) exceeded the performance of the foundation models.

Tables~\ref{tab:results3_gh2}, \ref{tab:results3_h2}, and \ref{tab:results3_gm2} summarise the results obtained using our proposed SARIMAX-based forecasting strategy. For each dataset, SARIMAX models are evaluated using either topological or statistical features as exogenous variables, with and without feature selection. The modelling framework first tests whether the inclusion of exogenous variables improves the baseline SARIMA performance. If improvement is obtained, feature selection is omitted to avoid unnecessary computational cost. Otherwise, feature selection is applied. To maintain clarity, intermediate results from this decision process are not reported; only the most effective configurations are presented.
\begin{table}[h!]
\centering
\resizebox{0.8\textwidth}{!}{%
\begin{tabular*}{\textwidth}{@{\extracolsep{\fill}}lllccc@{}}
	\toprule
	\textbf{Model} & \textbf{Feature Selection} & \textbf{Exogenous} & \textbf{MAE} & \textbf{MAPE} & \textbf{Cost} \\
	\midrule
	SARIMA   & None            & None         & 7.36 & 0.15 & 3    \\
	\midrule
	SARIMAX  & No             & Topological  & \textbf{3.14} & \textbf{0.07} & 80   \\
	SARIMAX  & Yes (RFE, PSO) & Topological  & --   & --   & --   \\
	\midrule
	SARIMAX  & No             & Statistical  & --   & --   & --   \\
	SARIMAX  & Yes (RFE, PSO) & Statistical  & 4.96 & 0.10 & 1016 \\
	\bottomrule
\end{tabular*}}
\caption{Performance of SARIMAX models on GH2 dataset using
different exogenous inputs and feature selection}
\label{tab:results3_gh2}
\end{table}

\begin{table}[h!]
\centering
\resizebox{0.8\textwidth}{!}{%
\begin{tabular*}{\textwidth}{@{\extracolsep{\fill}}lllccc@{}}
	\toprule
	\textbf{Model} & \textbf{Feature Selection} & \textbf{Exogenous} & \textbf{MAE} & \textbf{MAPE} & \textbf{Cost} \\
	\midrule
	SARIMA   & None            & None         & 3.88 & 0.10 & 7    \\
	\midrule
	SARIMAX  & No             & Topological  & --   & --   & --   \\
	SARIMAX  & Yes (RFE, PSO) & Topological  & \textbf{3.59} & \textbf{0.10} & 520  \\
	\midrule
	SARIMAX  & No             & Statistical  & --   & --   & --   \\
	SARIMAX  & Yes (RFE, PSO) & Statistical  & 5.35 & 0.13 & 1031 \\
	\bottomrule
\end{tabular*}}
\caption{Performance of SARIMAX models on H2 dataset using
different exogenous inputs and feature selection}
\label{tab:results3_h2}
\end{table}

\begin{table}[h!]
\centering
\resizebox{0.8\textwidth}{!}{%
\begin{tabular*}{\textwidth}{@{\extracolsep{\fill}}lllccc@{}}
	\toprule
	\textbf{Model} & \textbf{Feature Selection} & \textbf{Exogenous} & \textbf{MAE} & \textbf{MAPE} & \textbf{Cost} \\
	\midrule
	SARIMA   & None            & None         & 7.84 & 0.21 & 3    \\
	\midrule
	SARIMAX  & No             & Topological  & \textbf{4.74} & \textbf{0.14} & 106 \\
	SARIMAX  & Yes (RFE, PSO) & Topological  & --   & --   & --   \\
	\midrule
	SARIMAX  & No             & Statistical  & 8.61 & 0.24 & 157 \\
	SARIMAX  & Yes (RFE, PSO) & Statistical  & --   & --   & --   \\
	\bottomrule
\end{tabular*}}
\caption{Performance of SARIMAX models on GM2 dataset using
different exogenous inputs and feature selection}
\label{tab:results3_gm2}
\end{table}

We also evaluated the combined use of topological and statistical features as exogenous variables. This integrated approach did not lead to measurable performance gains over using the feature types individually and was therefore not included in the final modelling strategy.

Overall, the proposed approach consistently outperforms all benchmark models, achieving up to a 40\% reduction in both MAE and MAPE. The most significant improvements occur when topological features are incorporated without feature selection, particularly for the GH2 and GM2 datasets.

For the H2 dataset, however, feature selection is essential to improve predictive accuracy. The most effective configuration is a SARIMAX model with topological features selected via RFE and PSO. Because the feature-combination search is NP-hard, PSO was run 15 times, and the features most frequently selected were included in the final PSO feature set.

All final models satisfy standard diagnostic checks and were validated by domain experts. Computational cost remains manageable, and the SARIMAX model using topological features without feature selection provides a favourable balance between accuracy and efficiency.

\begin{figure}[h!]
    \centering
    \includegraphics[width=0.9\textwidth]{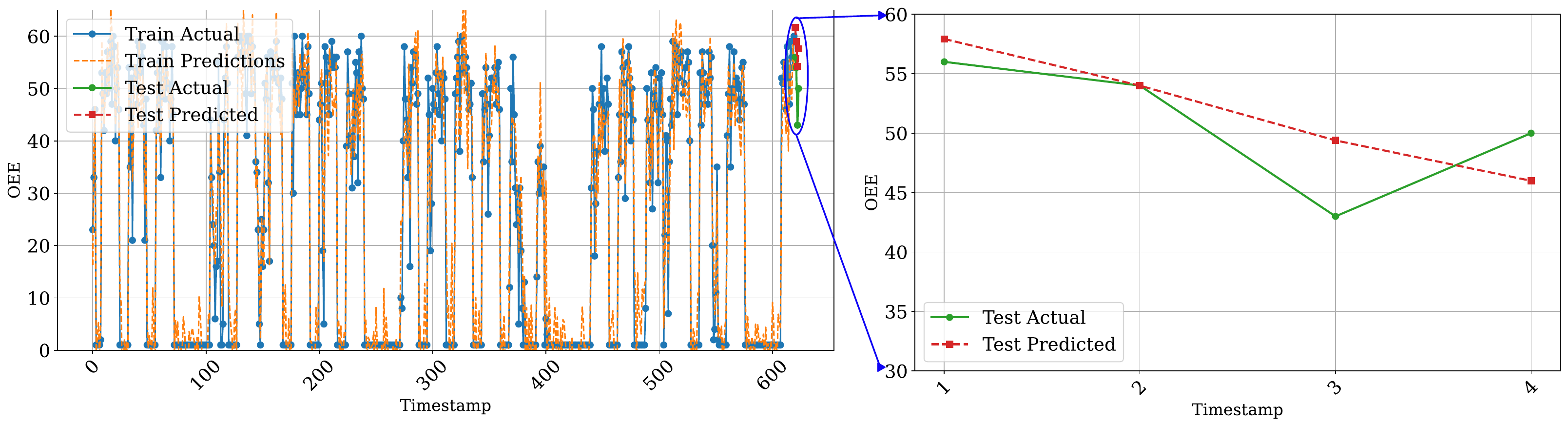}
    \caption{Forecasting result for dataset GH2}
    \label{fig:bestgh2}
\end{figure}

\begin{figure}[h!]
    \centering
    \includegraphics[width=0.9\textwidth]{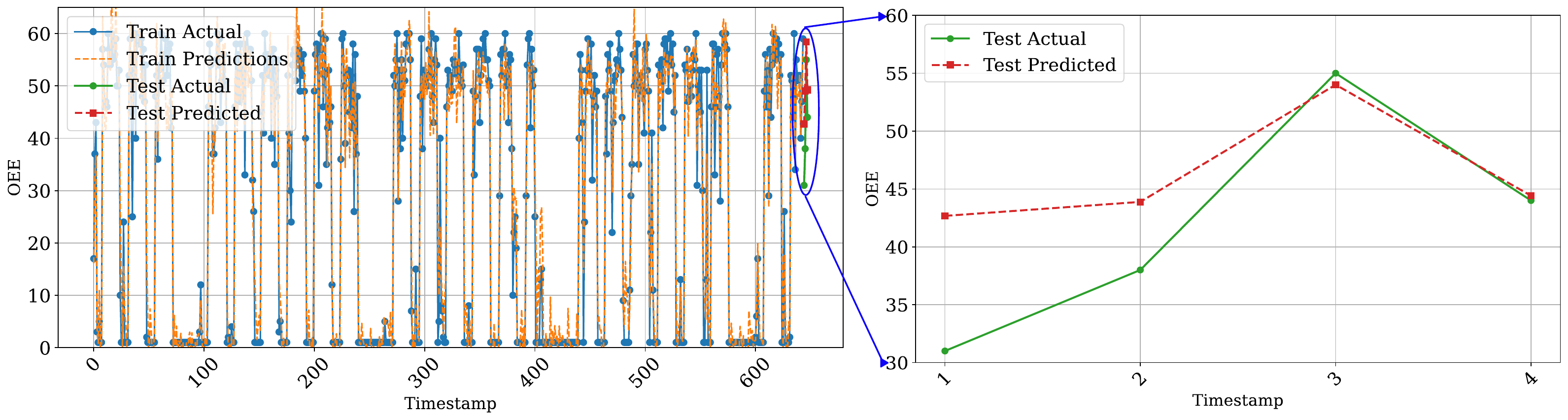}
    \caption{Forecasting result for dataset GM2}
    \label{fig:bestgm2}
\end{figure}

\begin{figure}[h!]
    \centering
    \includegraphics[width=0.9\textwidth]{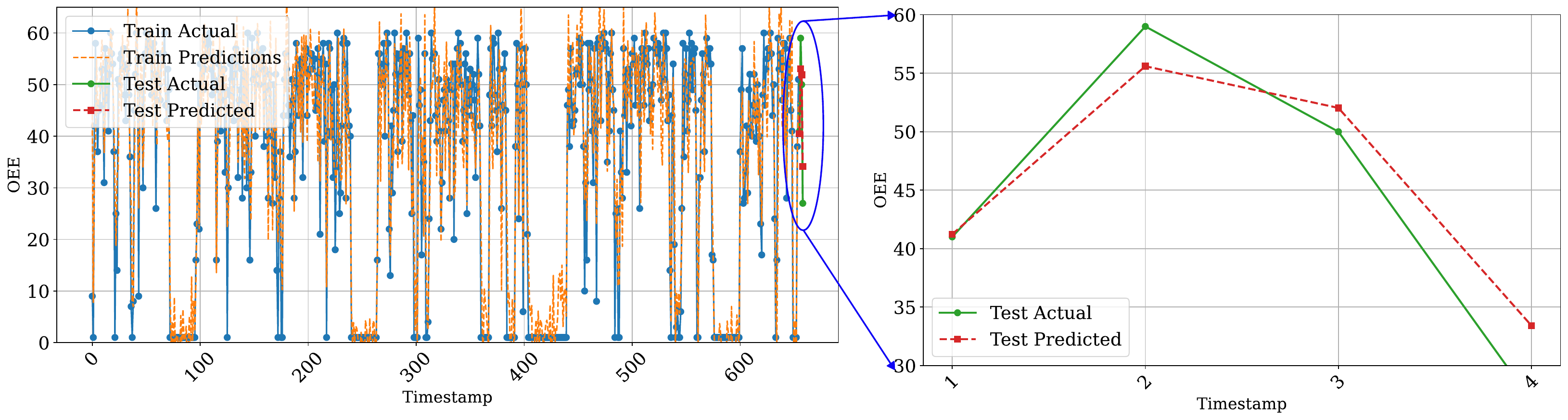}
    \caption{Forecasting result for dataset H2}
    \label{fig:besth2}
\end{figure}

Figures~\ref{fig:bestgh2}, \ref{fig:bestgm2} and \ref{fig:besth2} show four-hour-ahead forecasts. In each figure, the right portion corresponds to the test set, and the left to the training set, making overfitting easily detectable. The close alignment between training and testing curves indicates strong generalisation.

\begin{figure}[h!]
    \centering
    
    \begin{minipage}[b]{0.42\textwidth}
        \centering
        \includegraphics[width=\textwidth]{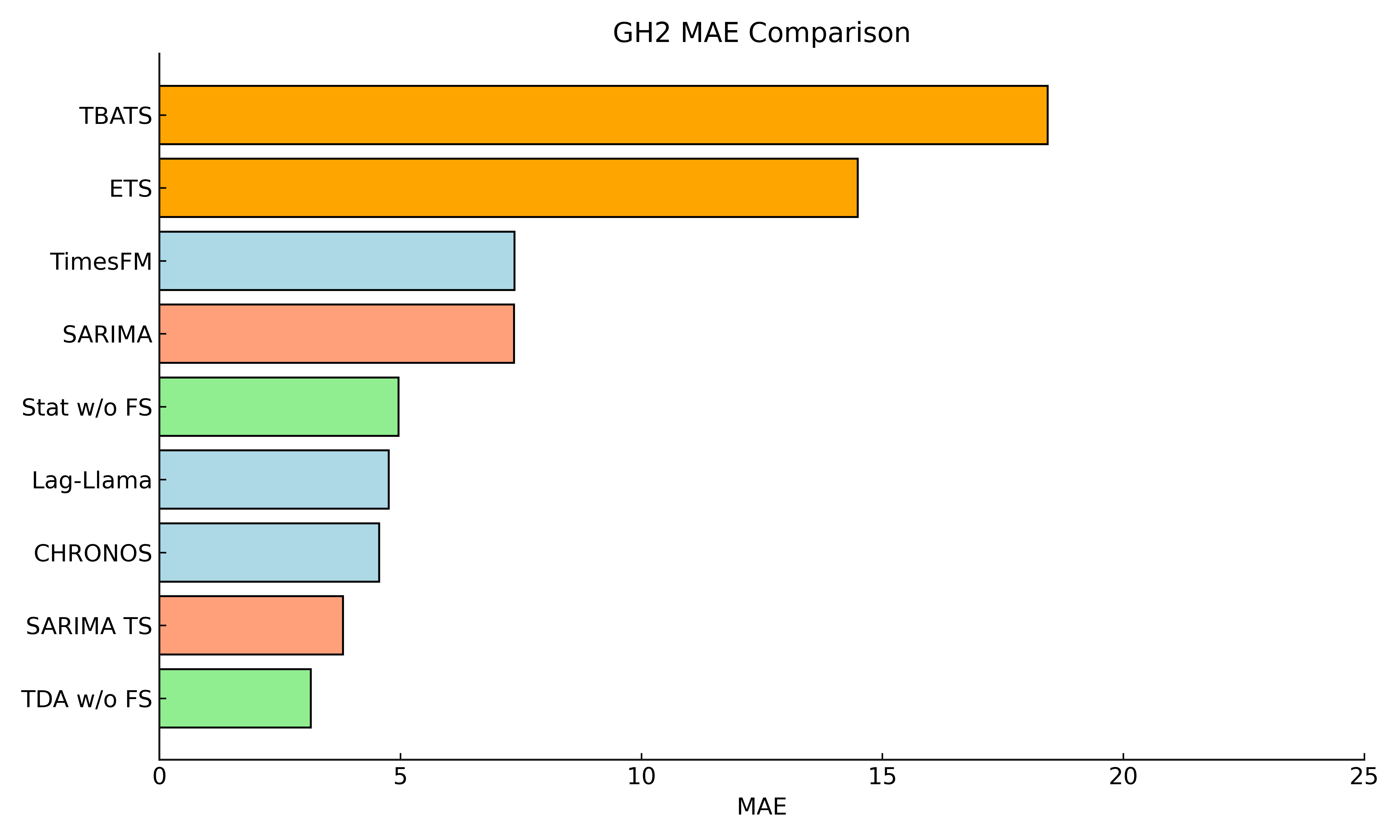}
        \caption*{(a) MAE for GH2 dataset}
    \end{minipage}
    \hfill
    \begin{minipage}[b]{0.42\textwidth}
        \centering
        \includegraphics[width=\textwidth]{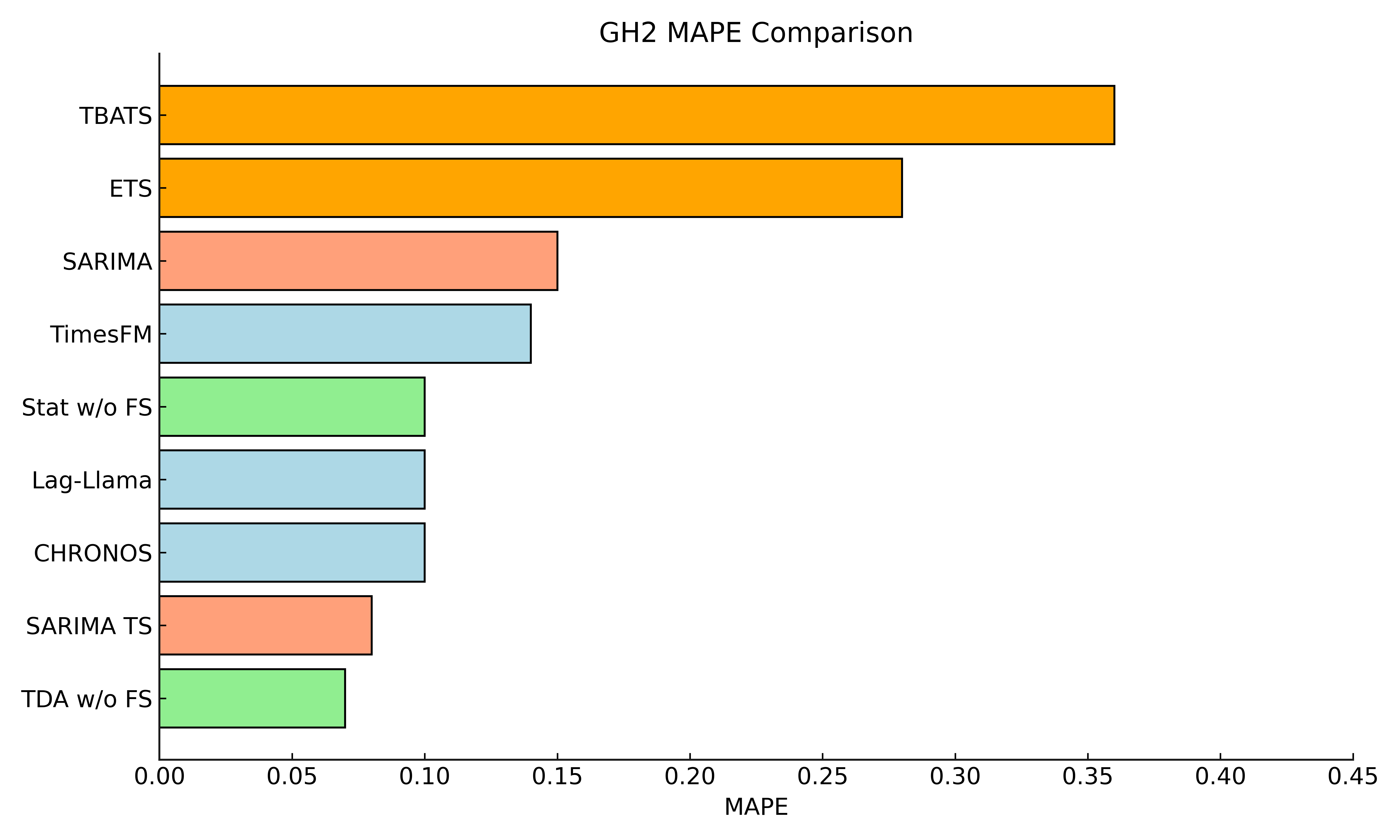}
        \caption*{(b) MAPE for GH2 dataset}
    \end{minipage}
    
    \vspace{1em}
    
    \begin{minipage}[b]{0.42\textwidth}
        \centering
        \includegraphics[width=\textwidth]{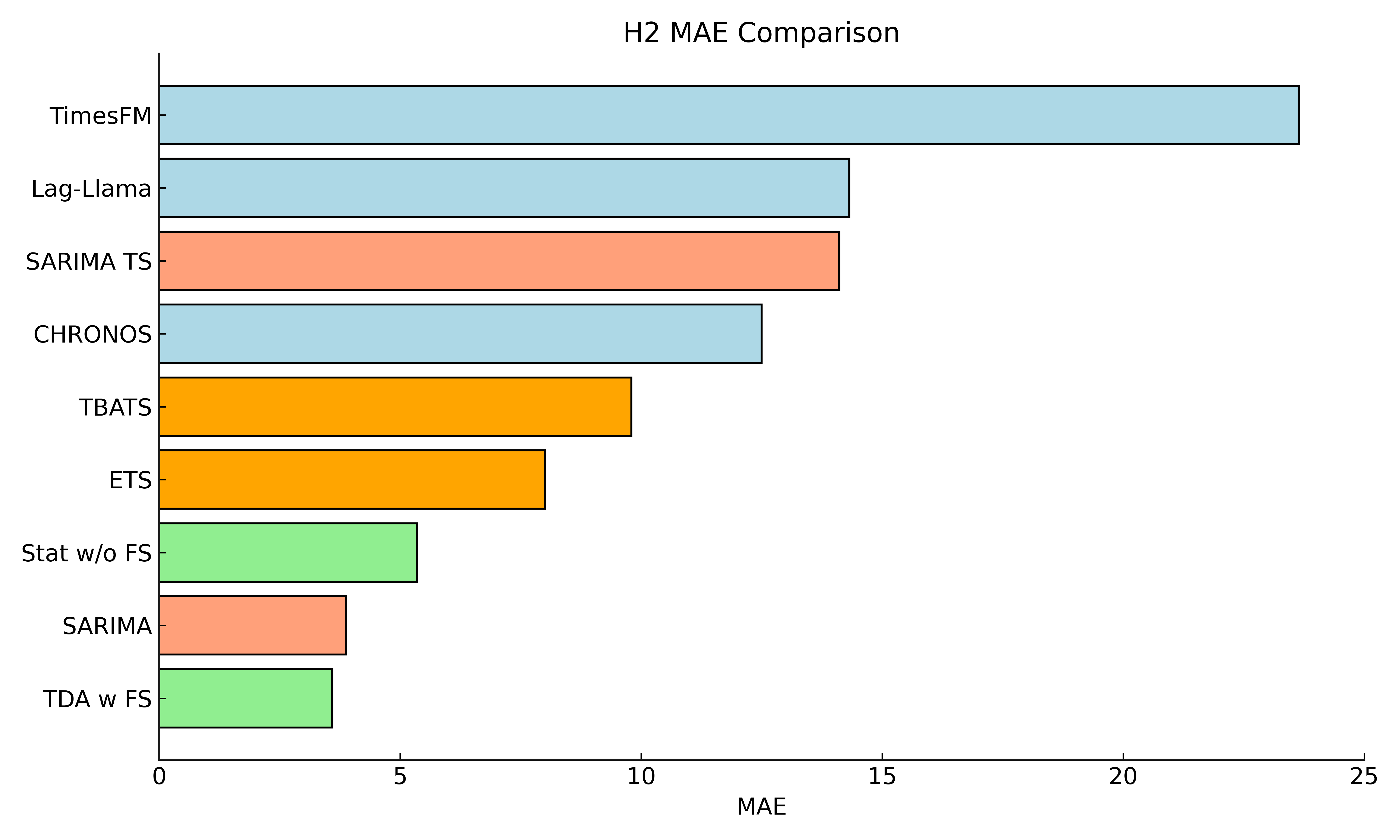}
        \caption*{(c) MAE for H2 dataset}
    \end{minipage}
    \hfill
    \begin{minipage}[b]{0.42\textwidth}
        \centering
        \includegraphics[width=\textwidth]{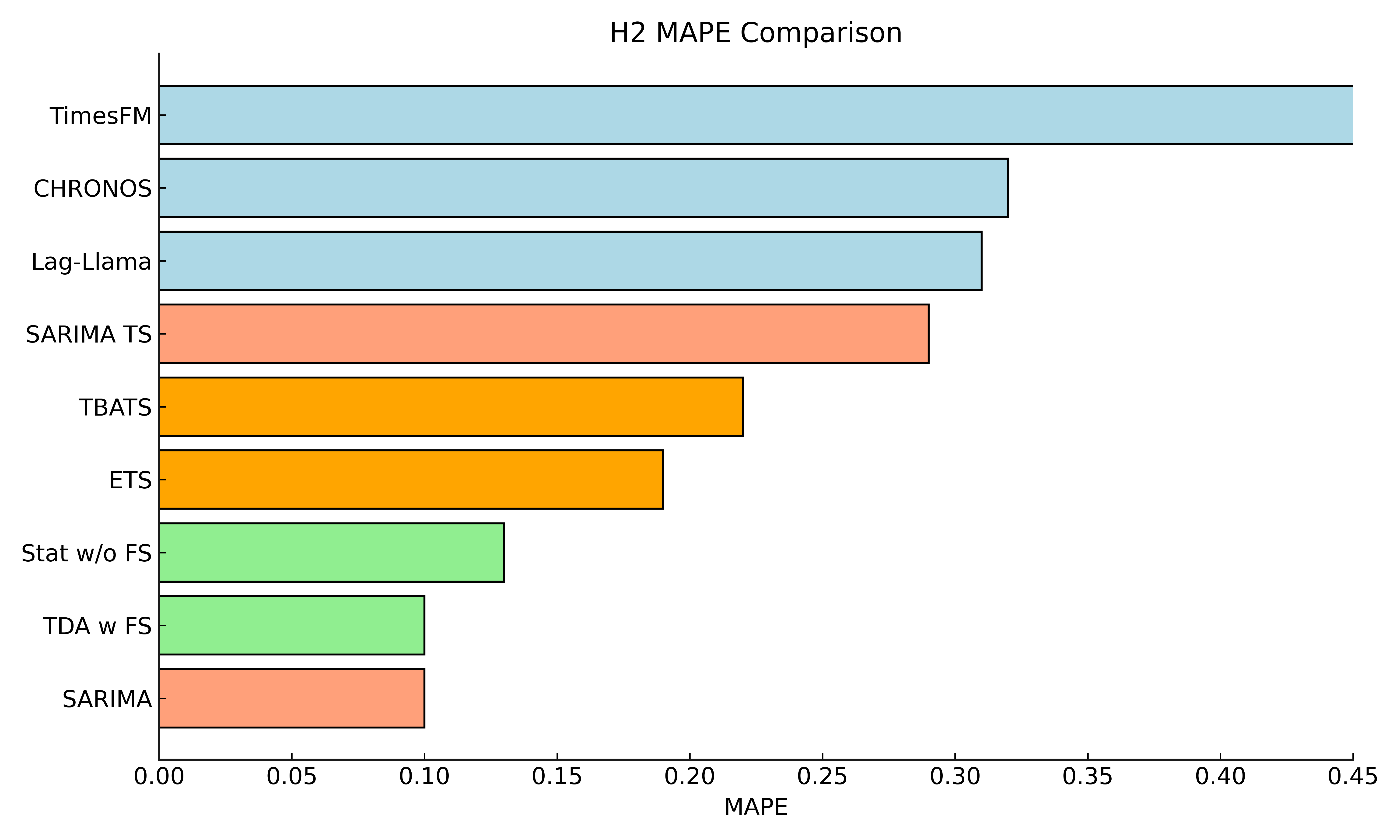}
        \caption*{(d) MAPE for H2 dataset}
    \end{minipage}
    
    \vspace{1em}
    
    \begin{minipage}[b]{0.42\textwidth}
        \centering
        \includegraphics[width=\textwidth]{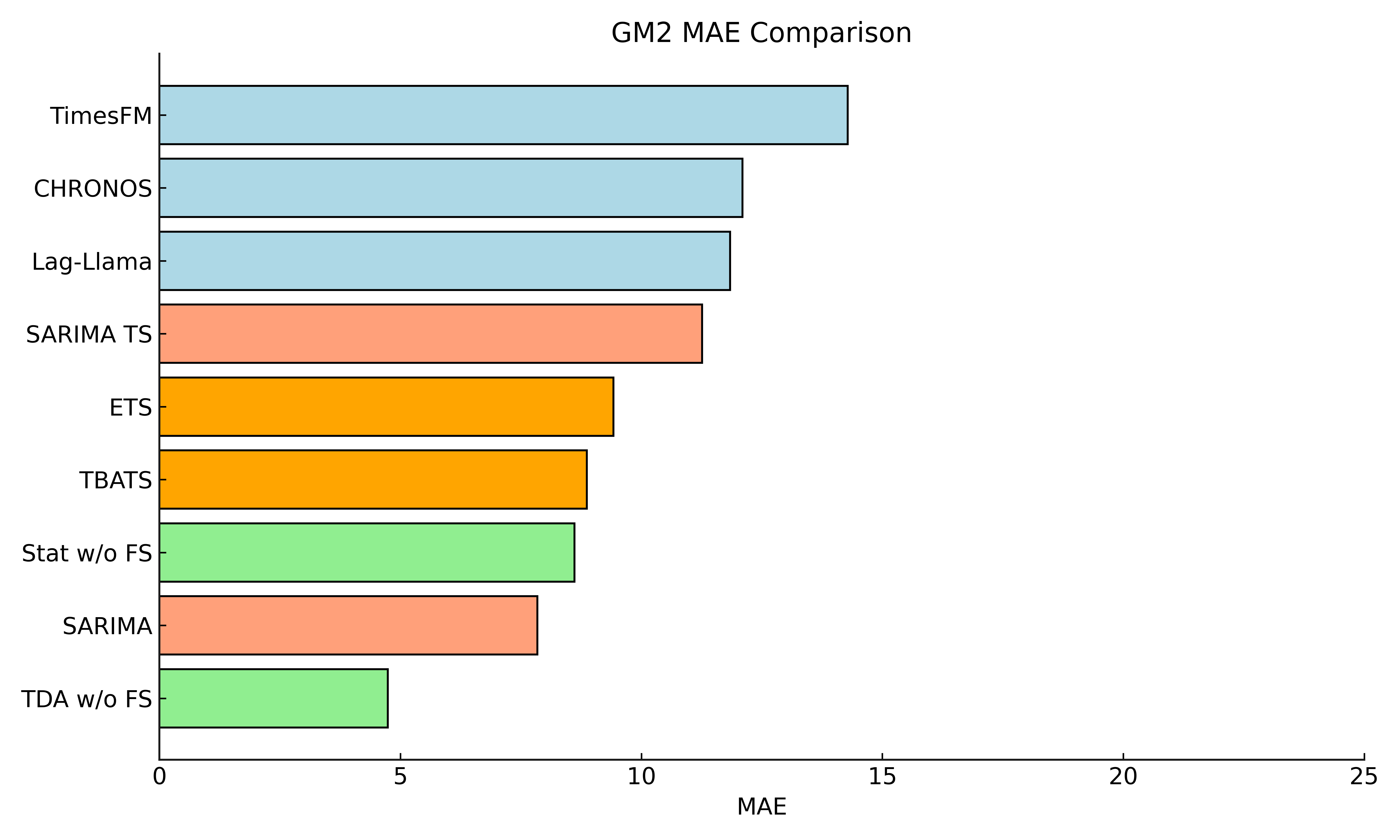}
        \caption*{(e) MAE for GM2 dataset}
    \end{minipage}
    \hfill
    \begin{minipage}[b]{0.42\textwidth}
        \centering
        \includegraphics[width=\textwidth]{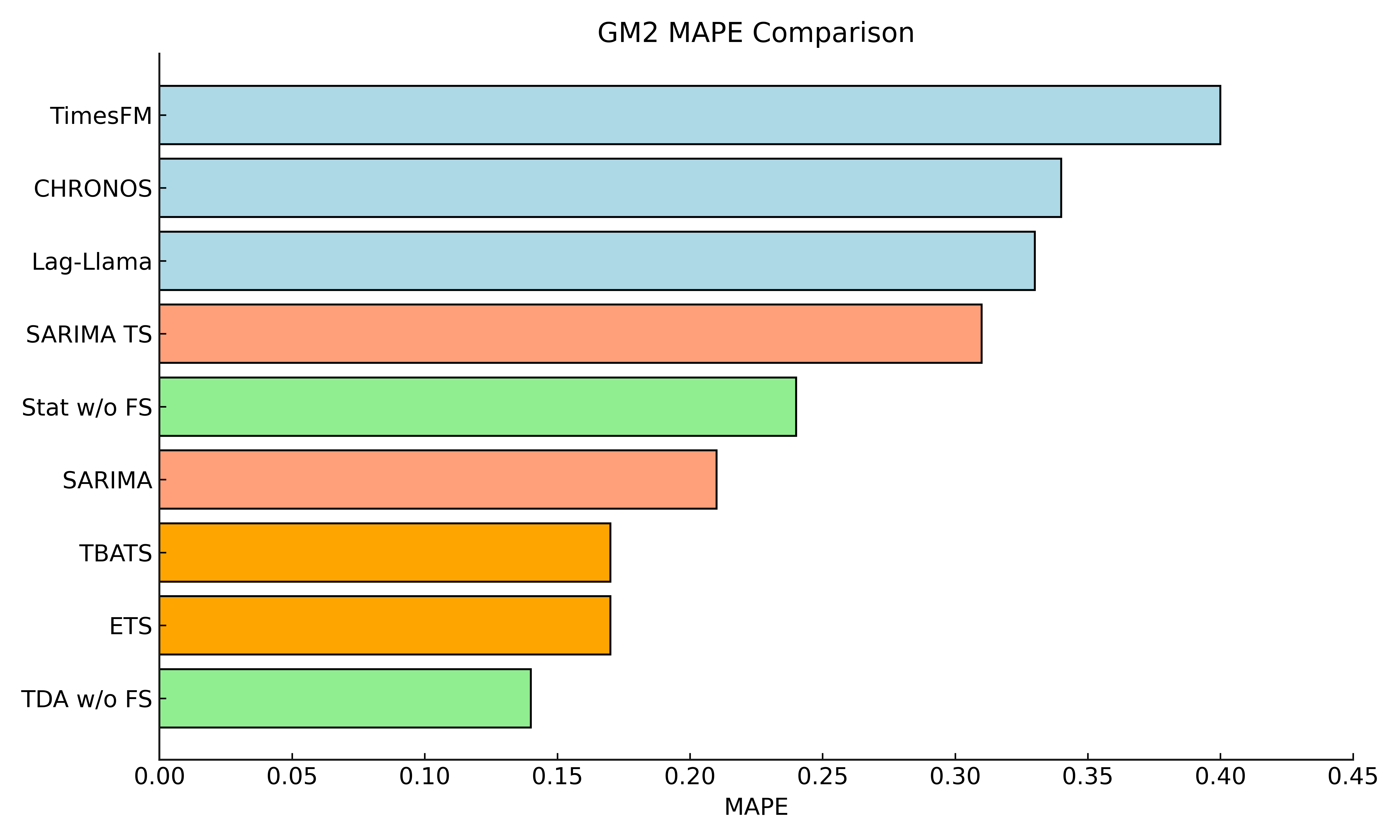}
        \caption*{(f) MAPE for GM2 dataset}
    \end{minipage}
    
    \caption{MAE and MAPE comparisons for GH2, H2 and GM2 datasets}
    \label{fig:comparison}
\end{figure}

Finally, Figure~\ref{fig:comparison} provides a comparative overview of forecasting performance across all models. Blue bars represent foundation models, green denotes our proposed model, orange denotes classical statistical models (ETS and TBATS), and red tones denote SARIMA models without exogenous variables. Across all datasets, the proposed approach with topological features consistently yields the lowest MAE and MAPE values.

\section{Industrial Case Study: GLN Production Plant Deployment}

\label{RWA}

OEE forecasting is most valuable when it is embedded directly into operational decision-support systems. To demonstrate the industrial applicability of the proposed methodology, this section presents its deployment in a manufacturing facility recognised within the GLN. The implementation targets proactive operational control by providing short-term OEE forecasts in a production environment characterised by high variability and frequent process interventions.

\begin{figure}[H]
	\centering
	\includegraphics[width=0.65\textwidth]{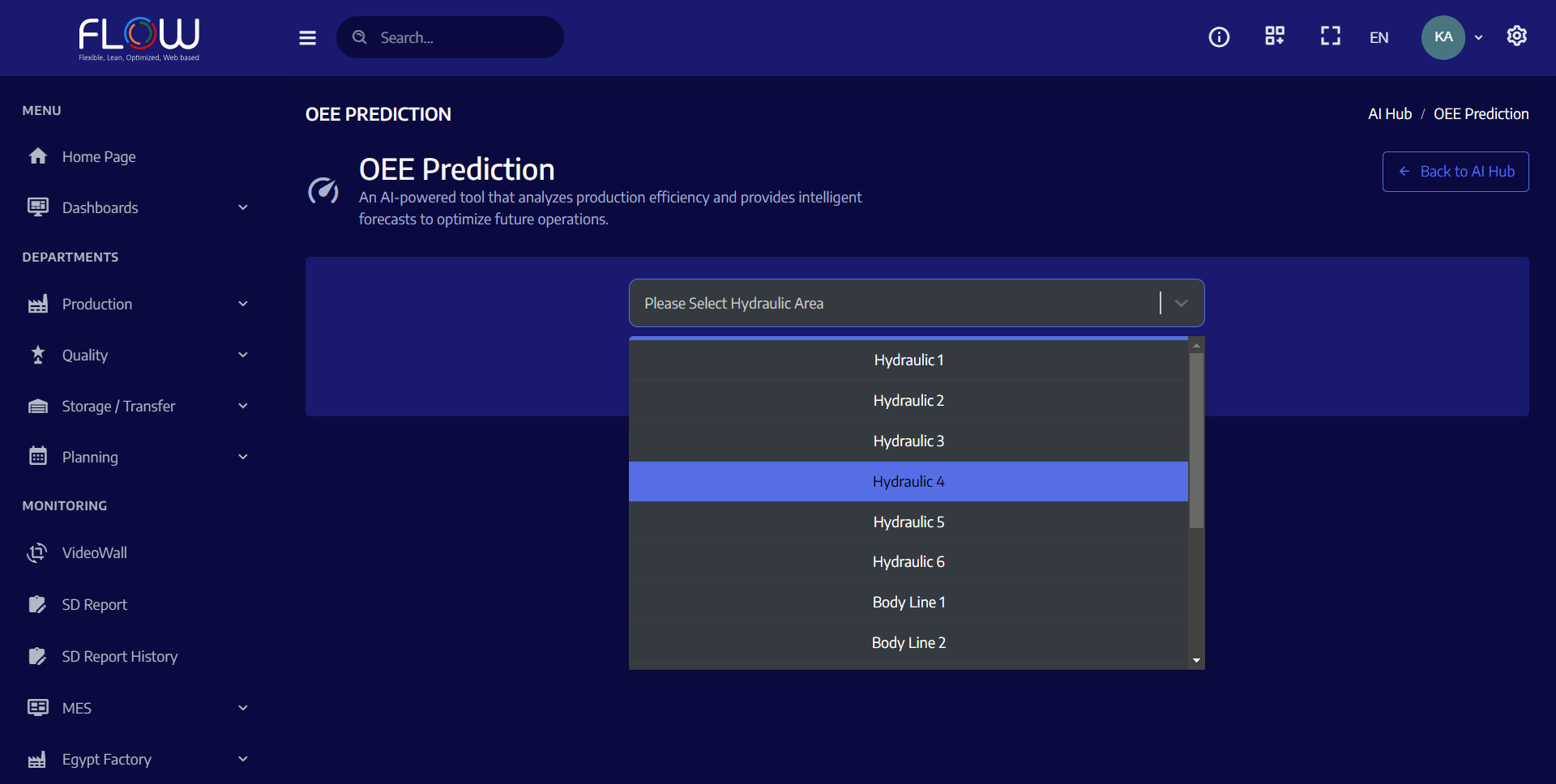}
	\caption{API interface displaying equipment list that can be selected to forecast the OEE values for next 4 hours.}
	\label{fig:api_interface}
\end{figure}

The proposed framework has been implemented through an Application Programming Interface (API) that continuously monitors and forecasts OEE values across 14 production assets. The system operates in real time and integrates seamlessly with existing manufacturing information systems. Figure~\ref{fig:api_interface} illustrates the main interface of the application, where users can select individual production areas and generate short-term OEE forecasts for upcoming operational horizons.

\begin{figure}[H]
	\centering
	\includegraphics[width=0.65\textwidth]{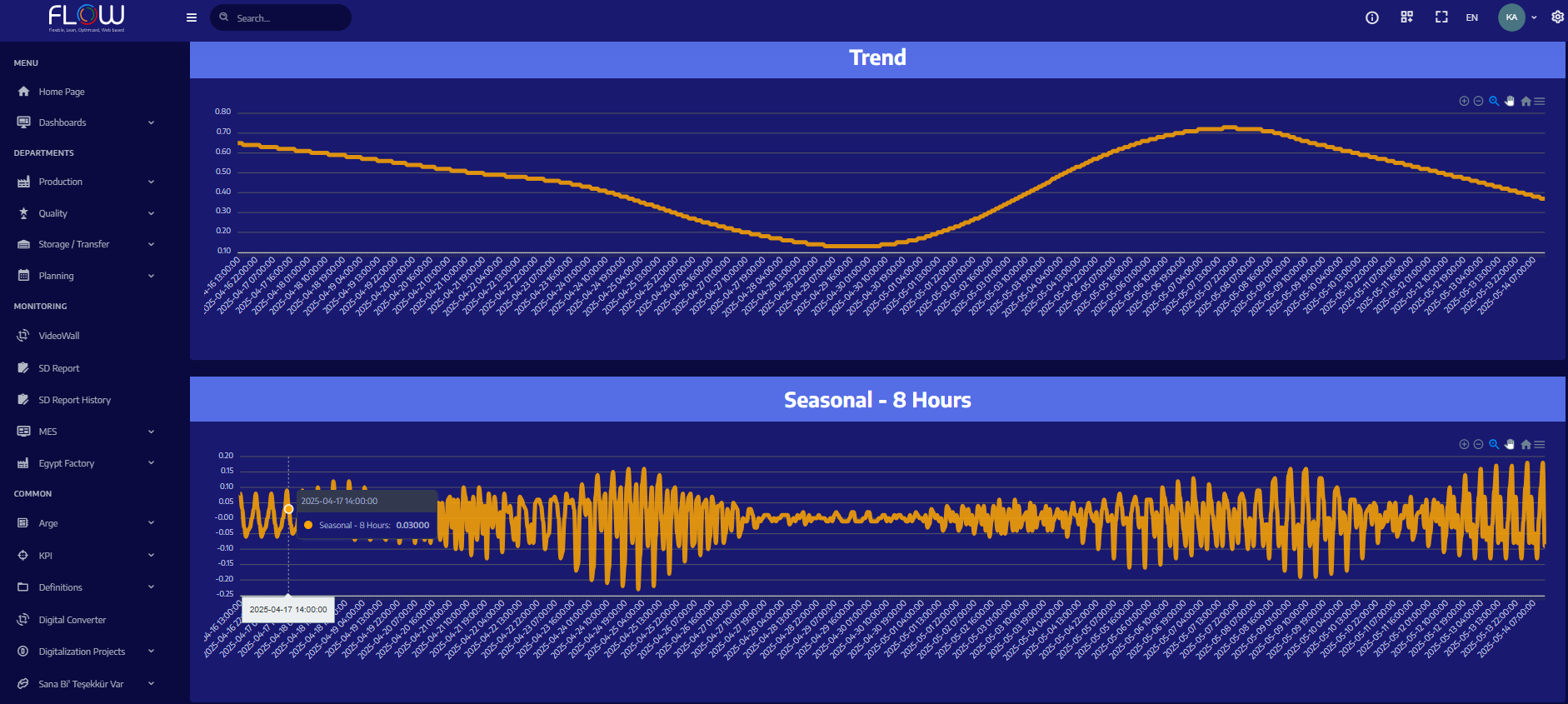}
	\caption{Visualization of trend and 8-hour seasonality components of OEE time series.}
	\label{fig:seasonality}
\end{figure}

The deployed system follows the decomposition-based forecasting strategy described in Section~\ref{sec:forecasting_models}. Incoming OEE time series are decomposed into trend, seasonal, and residual components. Multiple seasonalities inherent to manufacturing operations are explicitly identified, including 8-hour (shift-level), 24-hour (daily), and 168-hour (weekly) cycles. Figure~\ref{fig:seasonality} presents an example of the extracted trend and shift-level seasonal components. The trend component reflects long-term changes in equipment performance, allowing production managers to detect gradual efficiency degradation or improvement. Seasonal components reveal systematic operational patterns, such as efficiency drops during shift transitions or recurring night-shift inefficiencies. In Table~\ref{tab:oee_decomposition} several operational insights can be seen.

\begin{table}[h!]
\centering
\resizebox{\textwidth}{!}{%
\begin{tabular}{|c|p{5cm}|p{7.5cm}|}
\hline
\textbf{Seasonality} & \textbf{Insight Extracted} & \textbf{Real-World Management Application} \\
\hline
\textbf{8-hour (Shift-level)} & Identifies performance differences across shifts (e.g., morning vs. night). Reveals issues like operator fatigue or uneven workload. & Adjust shift scheduling, retrain operators, rebalance workload, or introduce automation in underperforming shifts. \\
\hline
\textbf{24-hour (Daily)} & Captures daily operational cycles including startup/shutdown inefficiencies or ambient condition effects. & Optimize warm-up/cool-down routines, realign maintenance timing, or dynamically adjust daily production targets. \\
\hline
\textbf{168-hour (Weekly)} & Highlights weekly rhythm: planned downtimes, production peaks, maintenance-induced performance drops. & Improve weekly planning, synchronize maintenance with low-demand periods, and stabilize batch production flows. \\
\hline

\end{tabular}%
}
\vspace{0.5em}
\caption{Operational insights derived from decomposition of OEE data (8h, 24h, 168h). These insights support decision-making in production environments.}
\label{tab:oee_decomposition}
\end{table}

\begin{figure}[H]
	\centering
	\includegraphics[width=0.65\textwidth]{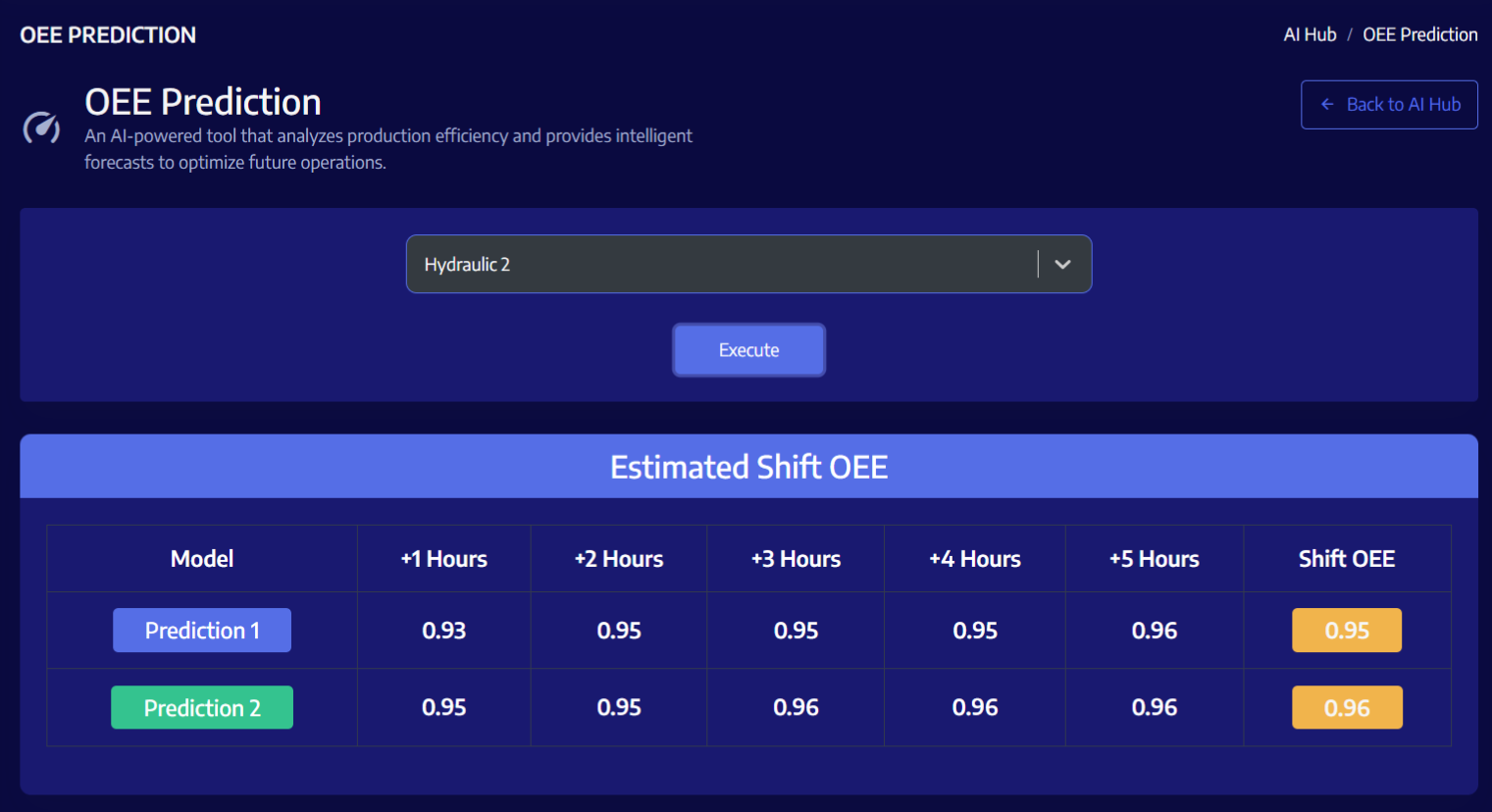}
	\caption{The forecast result on the API.}
	\label{fig:residual}
\end{figure}

The residual component captures short-term irregular fluctuations that are not explained by trend or seasonality. These residual deviations often correspond to micro-downtimes, unplanned slowdowns, or emerging quality issues. Continuous monitoring of the residual behaviour enables early detection of abnormal operating conditions and supports timely maintenance interventions. Figure~\ref{fig:residual} illustrates the forecasting outputs delivered through the API.

The forecasts enable maintenance and production teams to take preventive actions before performance degradation materialises. For instance, predicted drops in OEE can trigger proactive maintenance scheduling, workload redistribution, or process adjustments. In this way, the deployed system supports closed-loop, data-driven operational control rather than reactive performance assessment.

The industrial deployment of the proposed forecasting framework confirms its practical efficiency in a real manufacturing environment. Following its integration into daily operational workflows, the production plant achieved a measurable improvement in equipment performance. In particular, the mechanical production area recorded a 7.4\% increase in OEE within the evaluation period, indicating that short-term forecasting insights were successfully translated into operational actions.

Table~\ref{tab:deployment_kpi} summarises the key operational indicators observed before and after deployment of the forecasting system. Beyond the OEE improvement, the table highlights qualitative yet operationally meaningful changes in production management. Unplanned micro-downtimes were reduced as early warnings derived from residual behaviour enabled preventive interventions. Maintenance practices shifted from predominantly reactive responses toward proactive scheduling, supported by short-term forecasts. Moreover, shift-level visibility of OEE evolved from retrospective monitoring to real-time, forward-looking assessment, allowing production supervisors to anticipate performance deviations. Collectively, these changes elevated decision-support capabilities from descriptive reporting to predictive, knowledge-driven control.

\begin{table}[h]
\centering
\resizebox{0.85\textwidth}{!}{%
\begin{tabular}{lcc}
\toprule
\textbf{KPI} & \textbf{Before Deployment} & \textbf{After Deployment} \\
\midrule
Average OEE (\%)              & Baseline level            & +7.4\% improvement \\
Unplanned micro-downtimes     & Frequent                  & Reduced \\
Maintenance reaction time     & Reactive                  & Proactive \\
Shift-level OEE visibility    & Limited                   & Real-time forecasted \\
Decision-support capability   & Descriptive               & Predictive \\
\bottomrule
\end{tabular}}
\caption{Operational impact of the proposed forecasting framework before and after real-world deployment.}
\label{tab:deployment_kpi}
\end{table}

These outcomes demonstrate that integrating decomposition-based forecasting with topologically informed feature representations can yield tangible operational benefits in Industry~4.0 settings. The successful application in a GLN manufacturing facility further indicates that the proposed framework is scalable and transferable to other smart production systems seeking to enhance predictive maintenance, production planning and performance optimisation.

\section{Conclusion}

This study proposed a forecasting framework for short-term, highly volatile OEE time series by integrating statistical modelling with TDA. The approach addresses a key challenge in modern manufacturing analytics: extracting predictive knowledge from short, noisy, and dynamically evolving production data. By decomposing OEE signals into trend, seasonal, and residual components, the methodology enables targeted modelling of distinct temporal behaviours, leading to improved forecasting accuracy and robustness. This deployment directly operationalises the contributions outlined in Section~\ref{sec2}, demonstrating how the proposed knowledge-driven forecasting framework translates methodological advances into measurable improvements in manufacturing performance.

A central contribution of this work is the use of topological feature representations to characterise the intrinsic structure of OEE dynamics. Unlike conventional time-series features, TDA-based descriptors capture geometric and topological properties that remain informative under strong volatility and nonlinearity. Empirical results show that these features—particularly Heat Kernel representations—consistently emerge as dominant predictors when incorporated as exogenous variables within a SARIMAX framework, demonstrating their efficiency for knowledge-driven forecasting.

To ensure practical applicability, a two-stage feature selection strategy was employed, combining statistical significance-based elimination with Particle Swarm Optimisation guided by the Bayesian Information Criterion. This hybrid procedure substantially reduces dimensionality while preserving predictive performance, enhancing interpretability and computational efficiency—both essential for deployment in real production environments.

Compared with classical forecasting methods such as ARIMA and exponential smoothing, which often struggle with irregular and nonlinear OEE behaviour due to rigid structural assumptions, the proposed framework demonstrates superior adaptability and precision. Moreover, while recent transformer-based foundation models such as Lag-LlaMA offer strong generalisation capabilities, they typically rely on large-scale pretraining and operate as black-box predictors, limiting interpretability and control in short-term, data-scarce industrial settings. By unifying statistical decomposition, topological feature extraction, and model-based selection within a single forecasting pipeline, the proposed methodology provides a transparent and data-efficient alternative that aligns with Industry~4.0 principles and advances engineering informatics approaches for manufacturing performance analysis.

Finally, from an industrial perspective, the proposed method enables more reliable short-term OEE forecasting using limited data, supporting proactive maintenance planning, improved resource allocation, and data-driven operational decision-making in smart manufacturing systems. The practical relevance of this framework is further validated through its successful deployment in an industrial production environment, as detailed in Section~\ref{RWA}.

\section*{CRediT authorship contribution statement}

Korkut Anapa, İsmail Güzel, and Ceylan Yozgatlıgil contributed equally to the conceptualization, methodology, analysis, and writing of this manuscript. All authors reviewed and approved the final version of the manuscript.

\section*{Data availability}

The datasets generated and analyzed during the current study are available in the following GitHub repository:  
\url{https://github.com/korkutanapa/OEE_ARTICLE_STUDIES}

\section*{Code availability}

The code used for analysis and modeling in this study is available at:  
\url{https://github.com/korkutanapa/OEE_ARTICLE_STUDIES}

\section*{Declaration of competing interest}

The authors declare that they have no known competing financial interests or personal relationships that could have appeared to influence the work reported in this paper.

\section*{Declaration of generative AI and AI-assisted technologies in the writing process}

During the preparation of this work, the authors used ChatGPT (OpenAI) and Grammarly as a language support tool to improve English fluency and to check grammar and clarity. After using this tool, the authors carefully reviewed, edited, and verified the content to ensure accuracy and originality. The authors take full responsibility for the content of the publication.

\end{document}